\documentclass[prl,nofootinbib,11pt,tightenlines, floatfix,eqsecnum,superscriptaddress,showkeys]{revtex4-1}

\usepackage[utf8]{inputenc}          
\usepackage{graphicx,xcolor}         
\usepackage{array,dcolumn,longtable} 
\usepackage{amsmath,amssymb,slashed} 

\providecommand\lfstyle{}                   
\providecommand\romanup[1]{\text{#1}}       
\providecommand\greekup[1]{#1}              
\renewcommand\textsc{\MakeUppercase}

\usepackage{siunitx}                                  
\DeclareSIUnit{\fm}{\femto\metre}                     

\newcommand\rhic{\textsc{rhic}}

\newcommand\lhc{\textsc{lhc}}

\newcommand\phenix{\textsc{phenix}}

\newcommand\atlas{\textsc{atlas}}
\newcommand\cms{\textsc{cms}}

\newcommand\dabmod{\textsc{dab-mod}}
\newcommand\vusphydro{\text{v-\textsc{usp}hydro}}

\newcommand\qcd{\textsc{qcd}}

\newcommand\cgc{\textsc{cgc}}

\newcommand\fonll{\textsc{fonll}}

\declareslashed{}{\divslash}{0.08}{0}{\mathcal{D}}

\DeclareSIUnit{\fm}{\femto\metre}

\newcommand\proton{{\romanup{p}}}
\newcommand\deuteron{{\romanup{d}}}

\newcommand\muon{{\greekup{\mu}}}

\newcommand\Dmeson{{\romanup{D}}}

\newcommand\Dzero{{\Dmeson^0}}

\newcommand\pA{{\proton\romanup{A}}}
\newcommand\pPb{{\proton\romanup{Pb}}}
\newcommand\pAu{{\proton\romanup{Au}}}

\newcommand\dAu{{\deuteron\romanup{Au}}}
\newcommand\nucleusnucleus{{\romanup{AA}}}
\newcommand\PbPb{{\romanup{PbPb}}}

\newcommand\ArAr{{\romanup{ArAr}}}
\newcommand\XeXe{{\romanup{XeXe}}}
\newcommand\OO{{\romanup{OO}}}
\newcommand\HeAu{{\romanup{HeAu}}}
\newcommand\Oxigen{{\romanup{O}}}
\newcommand\U{{\romanup{U}}}
\newcommand\Xe{{\romanup{Xe}}}

\newcommand\Au{{\romanup{Au}}}
\newcommand\Ru{{\romanup{Ru}}}
\newcommand\Zr{{\romanup{Zr}}}
\newcommand\Ar{{\romanup{Ar}}}
\newcommand\snn[1][]{\sqrt{s_\text{NN}}\ifx\\#1\\\else=\SI{#1}{\TeV}\fi}

\newcommand\pt{p_\text{T}}

\newcommand\raa{R_\text{AA}}
\newcommand\vn[1]{v_{#1}}
\newcommand\vnn{\vn{n}}

\newcommand\Tfo{T_\text{FO}}

\begin{document}
\title{System size scan of $\Dmeson$ meson \textmd{$\raa$} and $\vnn$ using $\PbPb$, $\XeXe$, $\ArAr$, and $\OO$ collisions at \lhc}
\date{\today}

\author{Roland Katz}
\affiliation{SUBATECH, Universit\'e de Nantes, EMN, IN2P3/CNRS, 44307 Nantes, France}
\author{Caio A.~G.~Prado}
\affiliation{Institute of Particle Physics, Central China Normal University (CCNU), Wuhan, Hubei 430079, China}
\author{Jacquelyn Noronha-Hostler}
\affiliation{University of Illinois at Urbana-Champaign, Urbana, IL 61801, USA}
\author{Alexandre A.~P.~Suaide}
\affiliation{Instituto de F\'{i}sica, Universidade de S\~{a}o Paulo, C.P. 66318, 05315-970 S\~{a}o Paulo, SP, Brazil}

\begin{abstract}
Experimental measurements indicate no suppression (e.g.~$R_\pPb \sim 1$) but a surprisingly large $\Dmeson$ meson $v_2$ was measured in $\pPb$ collisions.  In order to understand these results we use Trento+\vusphydro+\dabmod\ to make predictions and propose a system size scan at the \lhc\ involving $^{208}\PbPb$, $^{129}\XeXe$, $^{40}\ArAr$, and $^{16}\OO$ collisions.  We find that the nuclear modification factor approaches unity as the system size is decreased, but nonetheless, in the 0--10\% most central collisions $\vn2\{2\}$ is roughly equivalent regardless of system size.  These results arise from a rather non-trivial interplay between the shrinking path length and the enhancement of eccentricities in small systems at high multiplicity.  Finally, we also find a surprising sensitivity of $\Dmeson$ mesons $\vn2\{2\}$ in 0--10\% at $\pt = 2$--\SI{10}{\GeV} to the slight deformation of $^{129}\Xe$ recently found at \lhc.
\end{abstract}
\maketitle

\noindent \textsl{1. Introduction.} The nature and properties of the smallest fluid known to humanity --- the Quark-Gluon Plasma --- has pushed the boundaries of our understanding of fluid dynamics.  Three significant signatures of the Quark-Gluon Plasma are collective flow, strangeness enhancement, and suppression of hard probes.  The first two signatures, collective flow and strangeness enhancement, have been measured in small asymmetric collisions such as $\pPb$, $\dAu$, and $^3\HeAu$~\cite{Chatrchyan:2013nka,Aaboud:2017acw,Aaboud:2017blb,Aad:2013fja,sirunyan:2018toe,Khachatryan:2014jra,Khachatryan:2015waa,Khachatryan:2015oea,Sirunyan:2017uyl,ABELEV:2013wsa,Abelev:2014mda,Adare:2013piz, Adare:2014keg,Aidala:2018mcw,Adare:2018toe,Adare:2015ctn,Aidala:2016vgl,Adare:2017wlc,Adare:2017rdq,Aidala:2017pup,Aidala:2017ajz,ALICE:2017jyt}.  Relativistic hydrodynamics manages to reproduce most flow observables in small systems well~\cite{Bozek:2011if,Bozek:2012gr,Bozek:2013ska,Bozek:2013uha,Kozlov:2014fqa,Zhou:2015iba,Zhao:2017rgg,Mantysaari:2017cni,Weller:2017tsr,Zhao:2017rgg}, although other scenarios that do not rely on relativistic hydrodynamics have been considered~\cite{Greif:2017bnr,Schenke:2016lrs,Mantysaari:2016ykx,Albacete:2017ajt}.  New experiments and measurements have been proposed in order to either confirm (or disprove) that relativistic hydrodynamics is the correct dynamical description in these tiny systems.  For instance, polarized beams~\cite{Bozek:2018xzy} and ultracentral deformed ion-ion collisions~\cite{Noronha-Hostler:2019ytn} both may distinguish between different scenarios in these light nuclei collisions.

The validity of event-by-event relativistic hydrodynamics remains to be studied in intermediate $\nucleusnucleus$ collisions across multiple types of ions.  It was recently proposed for \lhc\ energies to run a system size scan including $\ArAr$ and $\OO$ collisions~\cite{Citron:2018lsq} where a variety of hydrodynamic predictions have been made~\cite{Sievert:2019zjr,Lim:2018huo,Giannini:2019abh}.  More recently a system size scan has also been proposed for \rhic\ energies as well~\cite{Huang:2019tgz}.

While collective flow experiment/theory comparisons are quite intriguing in small systems, we still lack a fundamental understanding of why jet and heavy flavor suppression is not measured in small systems (e.g.~$R_{\pPb} \sim 1$ \cite{Adam:2015qda,Acharya:2019mno}). Predictions of $\XeXe$ $\raa$ have been done in~\cite{Zigic:2018ovr,Shi:2018vys} (although only with the assumption of spherical $^{129}\Xe$), quarkonium predictions for $\pA$ in~\cite{Zhang:2019dth} from the Color Glass Condensate (\cgc),  preliminary results for photons in $\pPb$, $\pAu$, $\dAu$, and $^3\HeAu$ collisions were shown in~\cite{Shen:2016egw}, and previous pPb calculations in a variety of scenarios can be found in \cite{Kang:2014hha,Xu:2015iha,Sharma:2009hn}.  However, we are not aware of any predictions available for intermediate systems such as $\ArAr$/$\OO$ nor any current azimuthal anisotropy predictions assuming event-by-event relativistic hydrodynamics as the medium that hard probes pass through. Meanwhile, the \cms\ collaboration has measured a significant $\Dmeson$ meson flow in $\pPb$ collisions~\cite{sirunyan:2017plt}, which is large but also somewhat suppressed compared to other identified particles. Additionally, the \atlas\ collaboration has measured a significant $\vn2$ from heavy flavor $\muon$'s in $\pPb$ collisions~\cite{ATLAS:2017zby}. It has also been suggested that the $R_{AA}$ to $v_2$ puzzle may be a good testing bed for initial conditions \cite{Djordjevic:2019tdu}.  Because $\Dmeson$ mesons are sensitive to equilibrium vs.~out-of-equilibrium dynamics (as shown in Fig.~10 from~\cite{Xu:2018gux}) combined with the significant $\vn2$ in small systems, they appear to be ideal candidates for understanding system size effects.

While in the \cgc\ scenario one can reproduce the experimentally measured heavy flavor $\vn2$~\cite{Zhang:2019dth}, no one has demonstrated in a initial condition+hydrodynamic+energy loss/Langevin scenario how such a large $\vn2$ in small systems is compatible with $\raa \rightarrow 1$.  Here we systematically study the effect of system size (by varying the colliding nuclei) on the nuclear modification factor, $\raa$, azimuthal anisotropies $\vnn\{2\}$, and multiparticle cumulants $\vn2\{4\}/\vn2\{2\}$.  To conduct this study we use Trento\,\cite{Moreland:2014oya}+\vusphydro\,\cite{Noronha-Hostler:2014dqa,Noronha-Hostler:2013gga}+\dabmod \,\cite{prado:2016szr} using the exact same soft sector backgrounds as in~\cite{Alba:2017hhe,Giacalone:2017dud,Sievert:2019zjr} and both the Langevin ``M\&T'' and the ``constant'' energy loss set up from~\cite{Katz:2019fkc} that works well compared to $\PbPb$ data respectively at low and high $p_T$.  We find that as the system size is shrunk $\raa\rightarrow 1$, however, the $\vnn\{2\}$'s have a rather non-trivial relationship with the system size that can only be understood with direct comparisons with the soft sector from~\cite{Sievert:2019zjr}.  Finally, we also find a non-trivial sensitivity to a deformed nucleons in intermediate $\pt = 2$ -- \SI{10}{\GeV} $\Dmeson$ meson $\vn2$ calculations, which is consistent with soft sector calculations~\cite{Giacalone:2017dud}.\\

\noindent \textsl{2. Model Description.} In this paper we couple 2D+1 event-by-event hydrodynamical backgrounds that fit the soft sector well to the heavy flavor code \dabmod~\cite{prado:2016szr,Katz:2019fkc} --- a modular Monte Carlo simulation package developed to study D and B mesons --- that samples heavy quarks using distributions from p\qcd\ \fonll\ calculations~\cite{Cacciari:1998it,Cacciari:2001td}.  Then, either a parametrized energy loss model that includes energy loss fluctuations (sampled from an underlying distribution) is used for the heavy quarks evolution or a relativistic Langevin model based on an input drag or diffusion coefficient. Once the decoupling temperature $T_d=160$ MeV is reached, hadronization follows via a hybrid fragmentation/coalescence model from which the final nuclear modification factor can be reconstructed.  In~\cite{Katz:2019fkc} it was shown for PbPb collisions that with the Langevin description using the purely collisional spatial diffusion coefficient model from~\cite{Moore:2004tg} ($D_s(2\pi T)=2.23$) one obtains a reasonable description of experimental data at low $\pt\lesssim 5$ -- \SI{6}{\GeV}, while for the high $\pt\gtrsim 5$ -- \SI{6}{\GeV} sector it was found that an energy loss model works best~\cite{Aaboud:2018bdg}.  However, the range of applicability for Langevin versus energy loss is not clear across system size and, therefore, we compare them across a somewhat wider range.  The initial conditions+hydrodynamical backgrounds are identical to those used in~\cite{Alba:2017hhe,Giacalone:2017dud,Sievert:2019zjr} where the Trento initial conditions used the parameters $p = 0$, $k = 1.6$, and $\sigma = \SI{0.51}{\fm}$, as established by a Bayesian analysis~\cite{bernhard:2016tnd}. The parameters of the viscous hydrodynamic code \vusphydro\ are $\tau_0 = \SI{0.6}{\fm}$, $\eta/s = 0.047$, $\Tfo = \SI{150}{\MeV}$, which have been shown to fit well compared to experimental data.  We test both a spherical $^{129}\Xe$ nucleus and a prolate one (per the parameterization of the deformed Wood-Saxon in~\cite{Moller:2015fba,Giacalone:2017dud}). 

One should note that in our model there is some ambiguity on the overall magnitude of $\raa$ in the absence of experimental data since we generally fix the scaling constant of the transport model using high $\pt$ $\raa$ in most central collisions.  Thus, it is possible that there may be a system size dependence to this constant, whereas we use here for all systems the value obtained in $\PbPb$ collisions. Furthermore, while we write $\vnn\{2\}$ to indicate that this is a two particle correlation --- obtained via the scalar product method for cumulants --- we also point out that one of these particles is a soft particle while the other is a heavy flavor hadron, as has been discussed extensively in~\cite{betz:2016ayq,noronha-hostler:2016eow, prado:2016szr, sirunyan:2017pan,Andres:2019eus}.\\

\begin{figure}[b]
    \centering
    \includegraphics[width=1.00\textwidth]{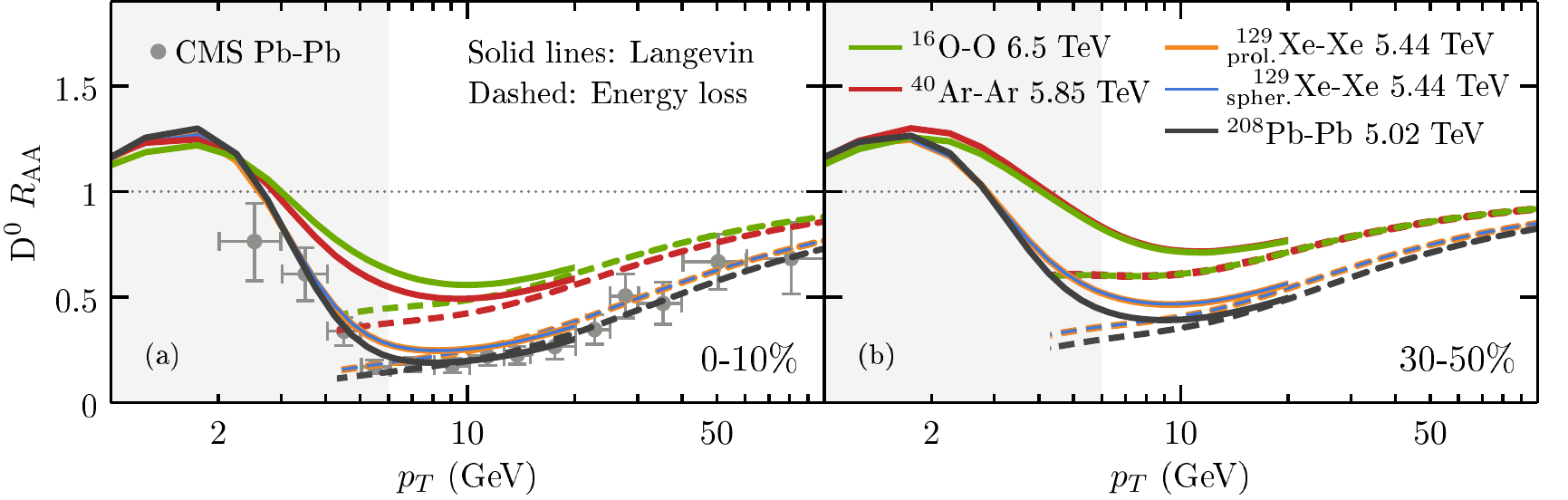}
    \caption{$\Dzero$ meson $\raa$ for $\PbPb$, $\XeXe$ with spherical and prolate initial nuclei, $\ArAr$, and $\OO$ collisions at the \lhc\ top energies in 0--10\% (a) and  30--50\% (b) centrality classes. The gray (white) area indicates the $\pt$ region where the Langevin (energy loss) description is the most relevant.}
    \label{fig:RAA}
\end{figure}

\begin{figure}[h]
    \centering
    \includegraphics[width=0.49\textwidth]{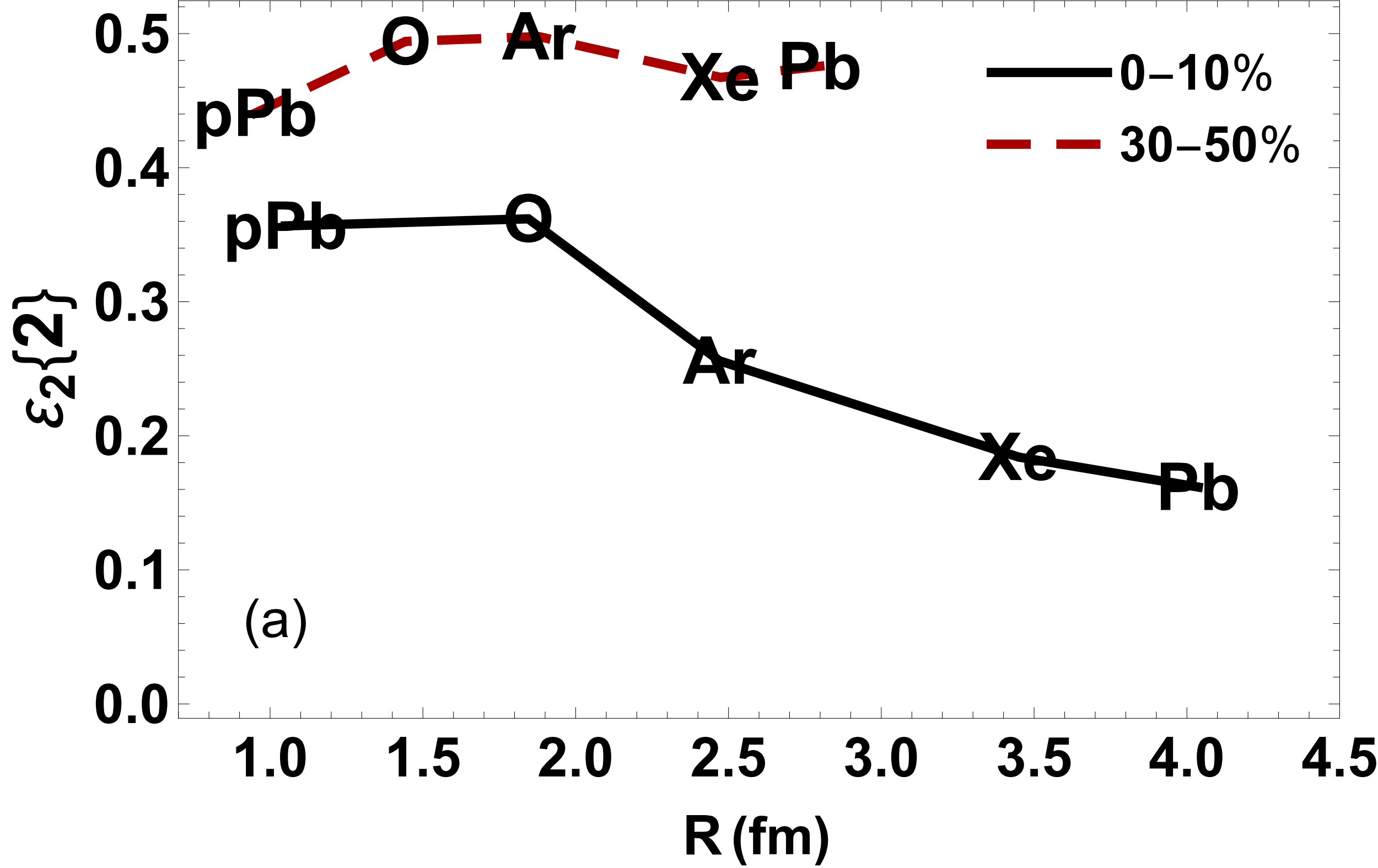}
\hspace{1mm}
    \includegraphics[width=0.49\textwidth]{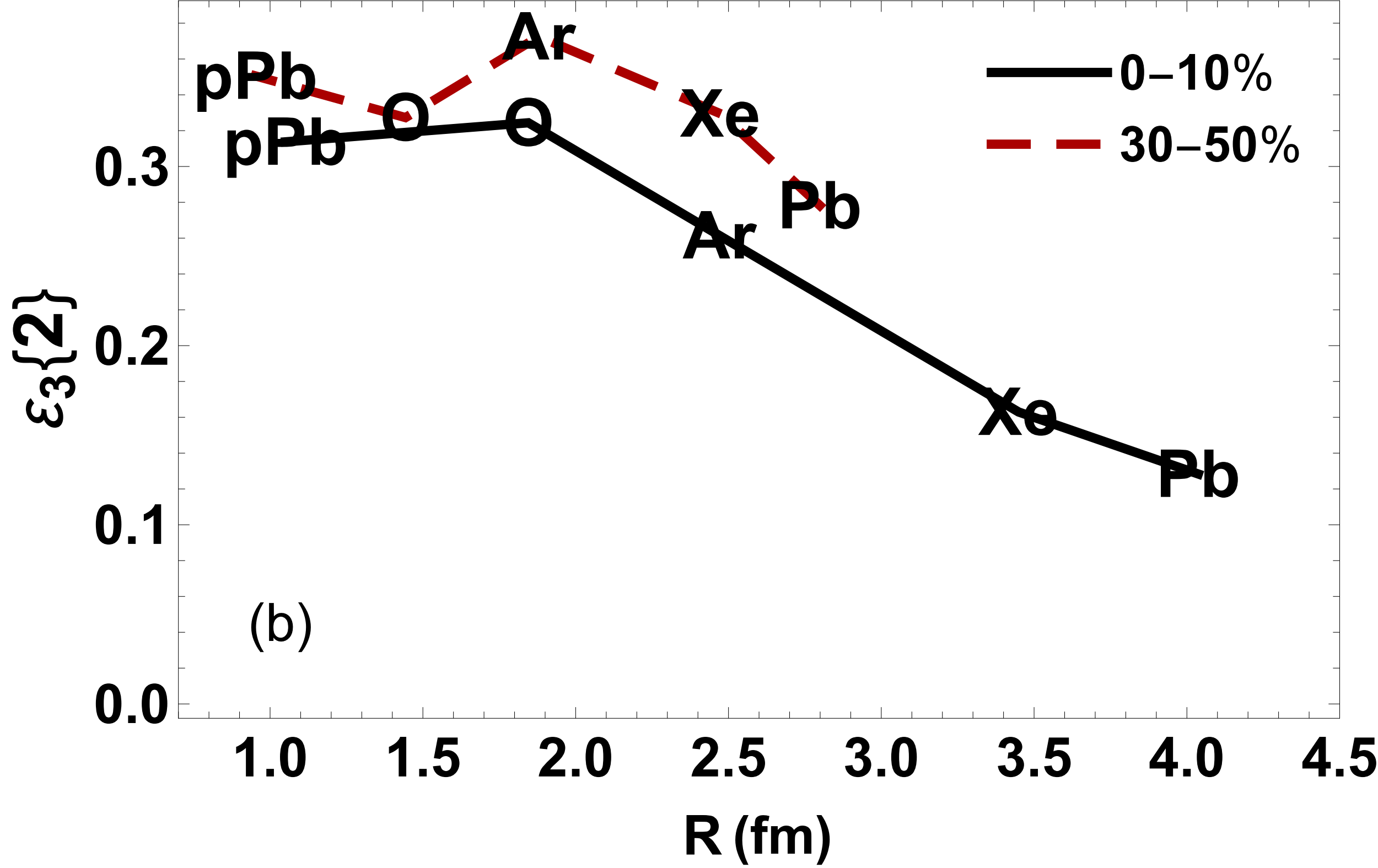}
    \caption{$\varepsilon_2\{2\}$ (a) and $\varepsilon_3\{2\}$ (b) versus radius for $\PbPb$, $\XeXe$, $\ArAr$, and $\OO$ collisions at the \lhc\ top energies in 0--10\% and  30--50\% centrality classes.}
    \label{fig:ecc}
\end{figure}

\noindent \textsl{3. Results.} In small systems it was found that the nuclear modification factor is consistent with unity within error bars~\cite{ALICE:2012mj,CMS:2013cka,Khachatryan:2016odn,ATLAS:2017dgr,Aaboud:2017cif,Bencedi:2016tks,Sett:2015pba,Todoroki:2017ngs}.  However, it is not clear how $\raa$ changes with system size as one moves towards small systems: does it smoothly increase as the size shrinks or does it suddenly jump to 1 at a certain critical size? Additionally, is the lack of light flavor jet suppression unique to asymmetric systems?  These questions are precisely investigated in Fig.~\ref{fig:RAA} where we show the $\raa$ of $\Dmeson$ mesons in 0--10\% and 30--50\% centrality classes.  There are a number of conclusions that can be drawn from these results. First, 0--10\% centralities are more sensitive to system size effects whereas for 30--50\% centralities one cannot see a distinguishable difference between $\ArAr$ and $\OO$ even though there is a clear difference in system size~\cite{Sievert:2019zjr}. We note that $\raa$ is insensitive to any effects of a deformed nucleus regardless of the centrality class. Second, it is clear that $\raa \rightarrow 1$ as the system size decreases, $(1-\raa)$ being roughly proportional to the system initial radius $\sim A^{1/3}$ from $\pt \gtrsim 4$ GeV, which implies that we expect a smooth decrease in the suppression of hard probes as one decreases the system size, eliminating the idea of a sharp critical system size within our framework. However, for the smallest system size of $\OO$ collisions there is still a rather important deviations from 1 at intermediate $\pt$.  Indeed, the  0--10\% centrality class can get a minimum close to $\raa \sim 0.5$ at its minimum whereas 30--50\% centrality class has $\raa \sim 0.7-0.8$. Extrapolating these results to $\pPb$ collisions - with radii of about two times smaller than $\OO$ collisions (see Fig.~\ref{fig:ecc}) - it is not obvious that the $\raa$ would reach unity enough to be consistent with $\pPb$ data. Finally, we find that Langevin and energy loss $\raa$ converge at high $\pt$ and the point of convergence occurs only at even higher $p_T$ for smaller systems. In contrast, at intermediate $\pt \sim 5$ -- 10 GeV the Langevin model predicts significantly less suppression than energy loss. We also point out that the difference between energy loss vs. Langevin in $\raa$ varies with system size due to different path length dependences. 


For the azimuthal anisotropies, it is important to understand that not only does the system size shrink but also the geometrical shape of the initial conditions change as well~\cite{Sievert:2019zjr}.  In Fig.~\ref{fig:ecc} we plot the radius of the initial conditions R versus the eccentricities $\varepsilon_n$ in the two centrality classes considered here: 0--10\% and  30--50\%. The systems coming from $\PbPb$, $\XeXe$, $\ArAr$, and $\OO$ central collisions have both significantly different sizes and significantly different eccentricities: as one decreases the radius, the eccentricities increase.  In contrast, mid-central collisions have roughly equivalent eccentricities and only vary in system size.  Thus, the mid-central collisions give us a better insight into pure system size effects.  However, one should caution that the $\Dmeson$ meson results from~\cite{sirunyan:2018toe} are measured in central $\pPb$ collisions so, to some extent, they may experience both varying system size and eccentricities compared to large $\nucleusnucleus$ collisions.

\vspace{5mm}

\begin{figure}[t]
    \centering
    \includegraphics[width=1.00\textwidth]{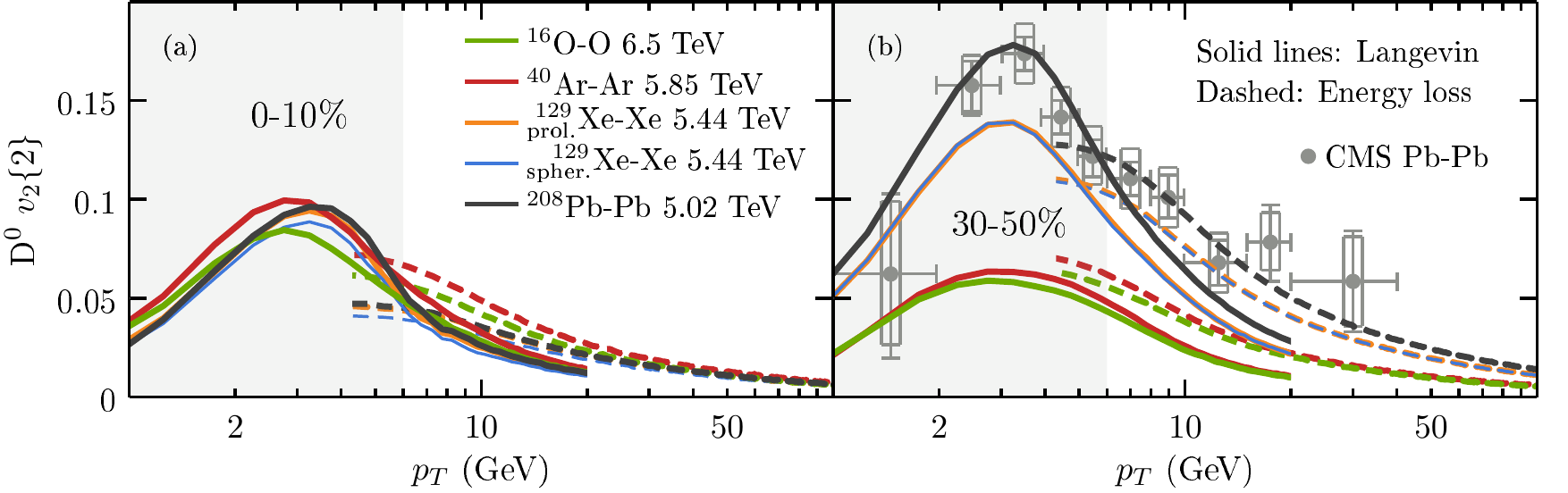}
    \caption{$\Dzero$ meson $\vn2\{2\}$ for $\PbPb$, $\XeXe$ with spherical and prolate initial nuclei, $\ArAr$, and $\OO$ collisions at the \lhc\ top energies in 0--10\% (a) and  30--50\% (b) centrality classes. The gray (white) area indicates the $\pt$ region where the Langevin (energy loss) description is the most relevant.}
    \label{fig:v2}
\end{figure}

The azimuthal anisotropies of hard probes can be useful too to study diffusion and energy loss~\cite{wang:2000fq,gyulassy:2000gk,Liao:2008dk,Jia:2011pi,Zapp:2013zya,betz:2014cza,Nahrgang:2014vza,Xu:2014tda,betz:2016ayq,noronha-hostler:2016eow, prado:2016szr,Xu:2017obm,Cao:2018ews,Xu:2018gux,Shi:2018vys}.   In Fig.~\ref{fig:v2} the elliptical azimuthal anisotropies are shown for $\Dmeson$ mesons at 0--10\% and 30--50\% centralities. We find that when we hold $\varepsilon_2 \sim \text{const.}$ in the 30--50\% centrality class the influence of the smaller system size plays a dramatic role in Fig.~\ref{fig:v2} for both descriptions of $\Dmeson$ dynamics where in small systems $\Dmeson$ mesons $\vn2$ is significantly suppressed across all $\pt$. Nevertheless, what is somewhat surprising is that $\vn2\{2\}(\pt)$ of $\OO$ collisions is roughly equivalent to $\vn2\{2\}(\pt)$ of $\ArAr$ collisions.

Now that we have shown that the system size suppresses $\Dmeson$ meson $\vn2$ when the eccentricities are held fixed, we can understand the results in the 0--10\% centrality class in Fig.~\ref{fig:v2} where the $\vn2$ is roughly equivalent regardless of system size. Additionally, the $\vn2$ in $\ArAr$ and $\OO$ collisions is larger in central collisions than in mid-central collisions.  Returning to Fig.~\ref{fig:ecc} we know that for central collisions as the system size decreases the eccentricities increase. Thus, there are now two competing factors that can contribute to the final $\vn2$: a suppression effect from decreasing the system size and an enhancement effect from increasing eccentricities.  In Fig.~\ref{fig:v2} when we see that all curves are very similar in 0--10\% centrality it simply is because these two competing effects roughly cancel each other out.  This implies that in the \cms\ $\pPb$ $\Dmeson$ mesons data likely there is a large enough eccentricity such that $\vn2$ does not vanish completely due to shrinking system size (although it may be that some initial flow could also influence $\Dmeson$ meson $\vn2$ \cite{Sun:2019fud}, we have not yet explored this possibility).

\begin{figure}[t]
    \centering
    \includegraphics[width=0.53\textwidth]{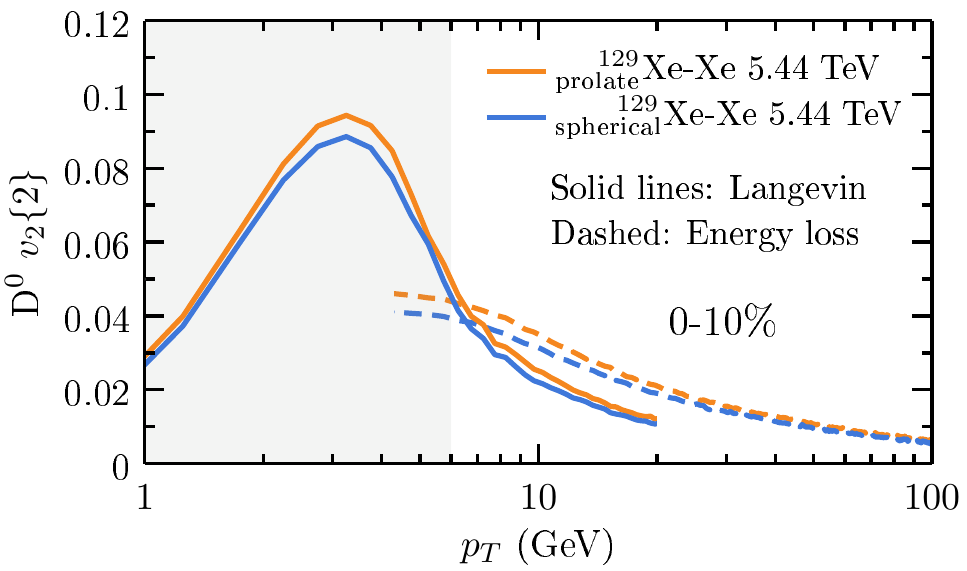}
    \caption{$\Dzero$ meson $\vn2\{2\}$ for $\XeXe$ collisions with spherical and prolate initial nuclei at the \lhc\ top energies in the 0--10\% centrality class. The gray (white) area indicates the $\pt$ region where the Langevin (energy loss) description is the most relevant.}
    \label{fig:v2XeXe}
\end{figure}

Another interesting consequence from Fig.~\ref{fig:v2} in 0--10\% centrality (shown explicitly Fig.~\ref{fig:v2XeXe}) is that the $\vn2$ between $\pt=2$ -- \SI{5}{\GeV} for Langevin and up to $\pt=$ \SI{10}{\GeV} for energy loss show some sensitivity to the deformation present in the $^{129}\Xe$ nucleus.  Using a prolate nucleus we find that there is an enhancement in this regime compared to a spherical nucleus.  This is a surprising result since 0--10\% is quite a wide centrality bin whereas in the soft sector the large deformation effects come from primarily ultracentral collisions \cite{Broniowski:2013dia,Adamczyk:2015obl,wang:2014qxa,Moreland:2014oya,Goldschmidt:2015kpa,Giacalone:2017dud,Rybczynski:2017nrx,Schenke:2019ruo,CMS:2018jmx,Acharya:2018ihu,ATLAS:2018iom}. Finally, although we expect a larger $v_2$ at a fixed $\pt\gtrsim 6$ GeV for the energy loss compared to Langevin in all systems, we find that they are both influenced by system size in a roughly equivalent way.

\begin{figure}[t]
    \centering
    \includegraphics[width=1.00\textwidth]{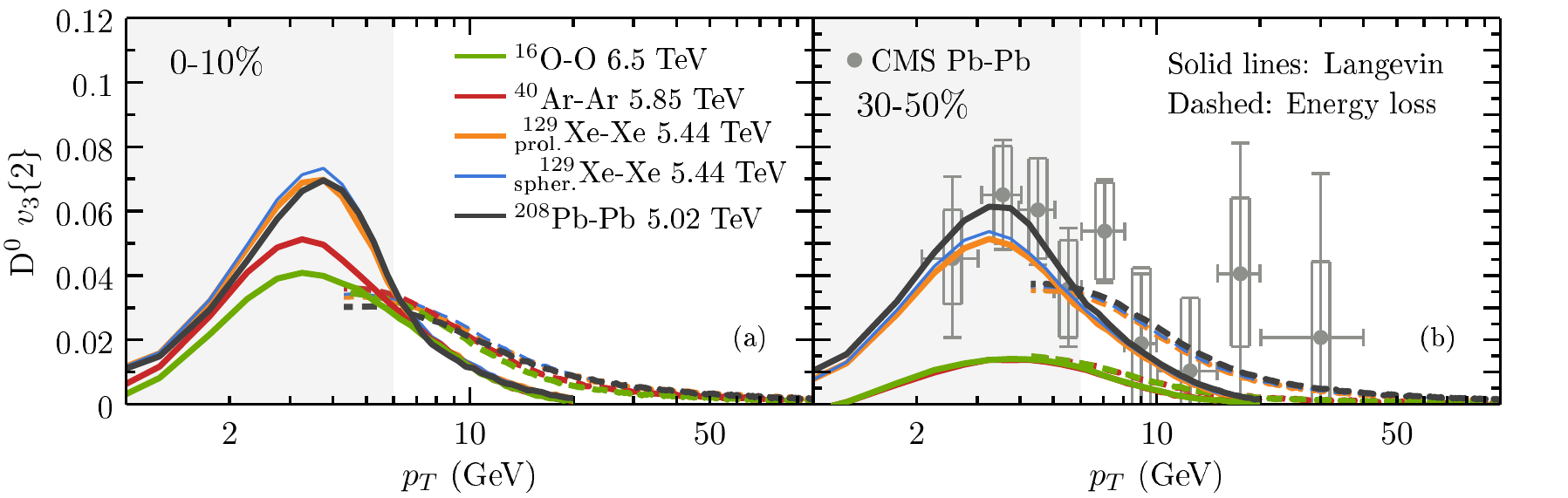}
    \caption{$\Dzero$ meson $\vn2\{2\}$ for $\PbPb$, $\XeXe$ with spherical and prolate initial nuclei, $\ArAr$, and $\OO$ collisions at the \lhc\ top energies in 0--10\% (a) and  30--50\% (b) centrality classes. The gray (white) area indicates the $\pt$ region where the Langevin (energy loss) description is the most relevant.}
    \label{fig:v3}
\end{figure}

We also explore these effects on $\vn3\{2\}(\pt)$ in Fig.~\ref{fig:v3} and find that $\vn3$ is more sensitive to system size effects i.e.~$\vn3$ is more consistently suppressed in small systems in both centrality classes, even when there is a significant increase in $\varepsilon_3$. The one exception to this is 0--10\% centrality for energy loss where all systems are nearly identical. Additionally, we find that the approximate universality of $\vn3\{2\}(\pt)$ across centralities in $\PbPb$ collisions (see also~\cite{sirunyan:2017plt,Nahrgang:2014vza}) is not observed in smaller systems. This approximate universality can then be explained by a balance between the variations in path length and eccentricity with centrality. The different response of $\vn2$ and $\vn3$ to system size dependence is quite interesting and may be helpful to constrain certain model parameters that could be considered in future studies.

\begin{figure}[t]
    \centering
    \includegraphics[width=0.49\textwidth]{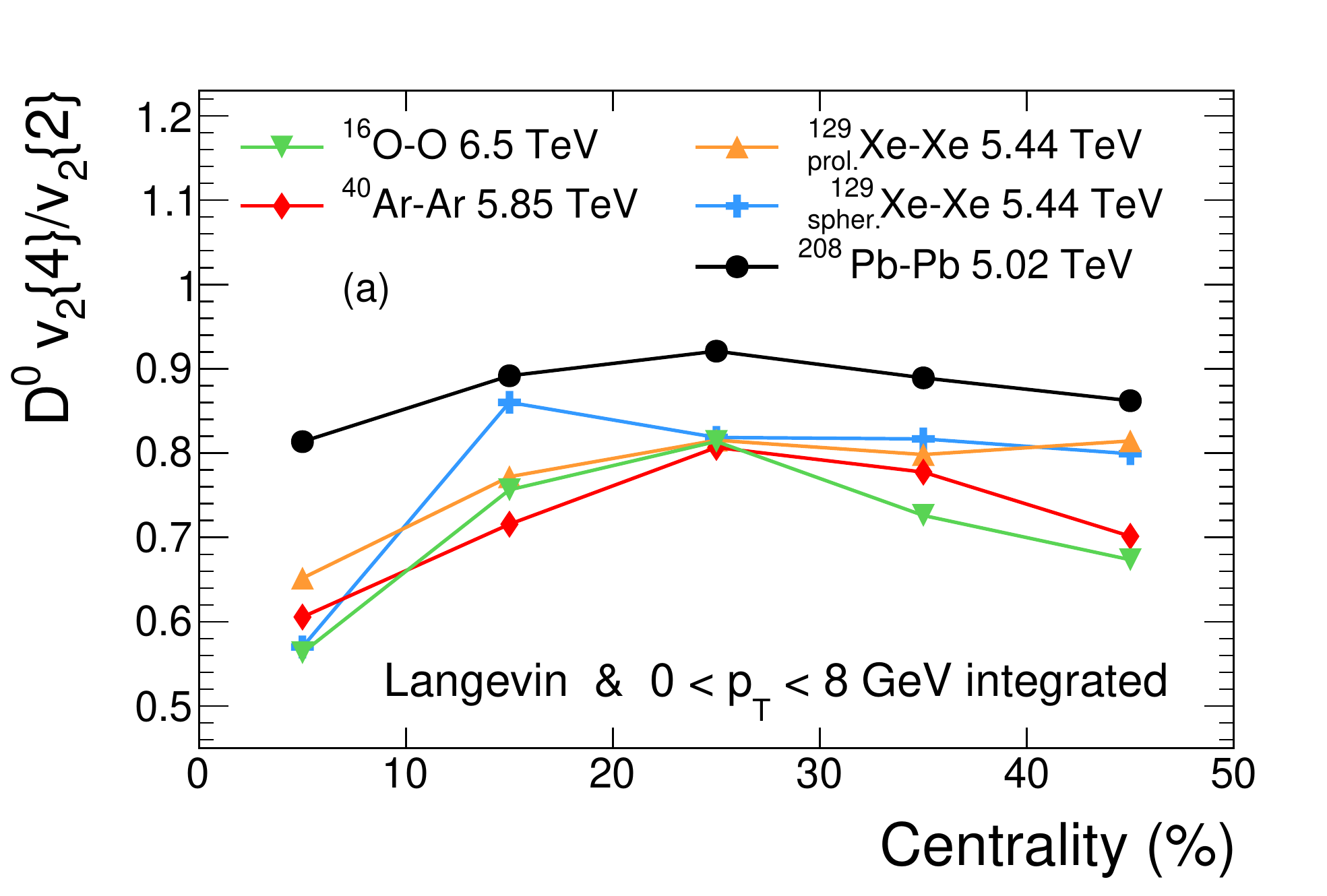}
    \includegraphics[width=0.49\textwidth]{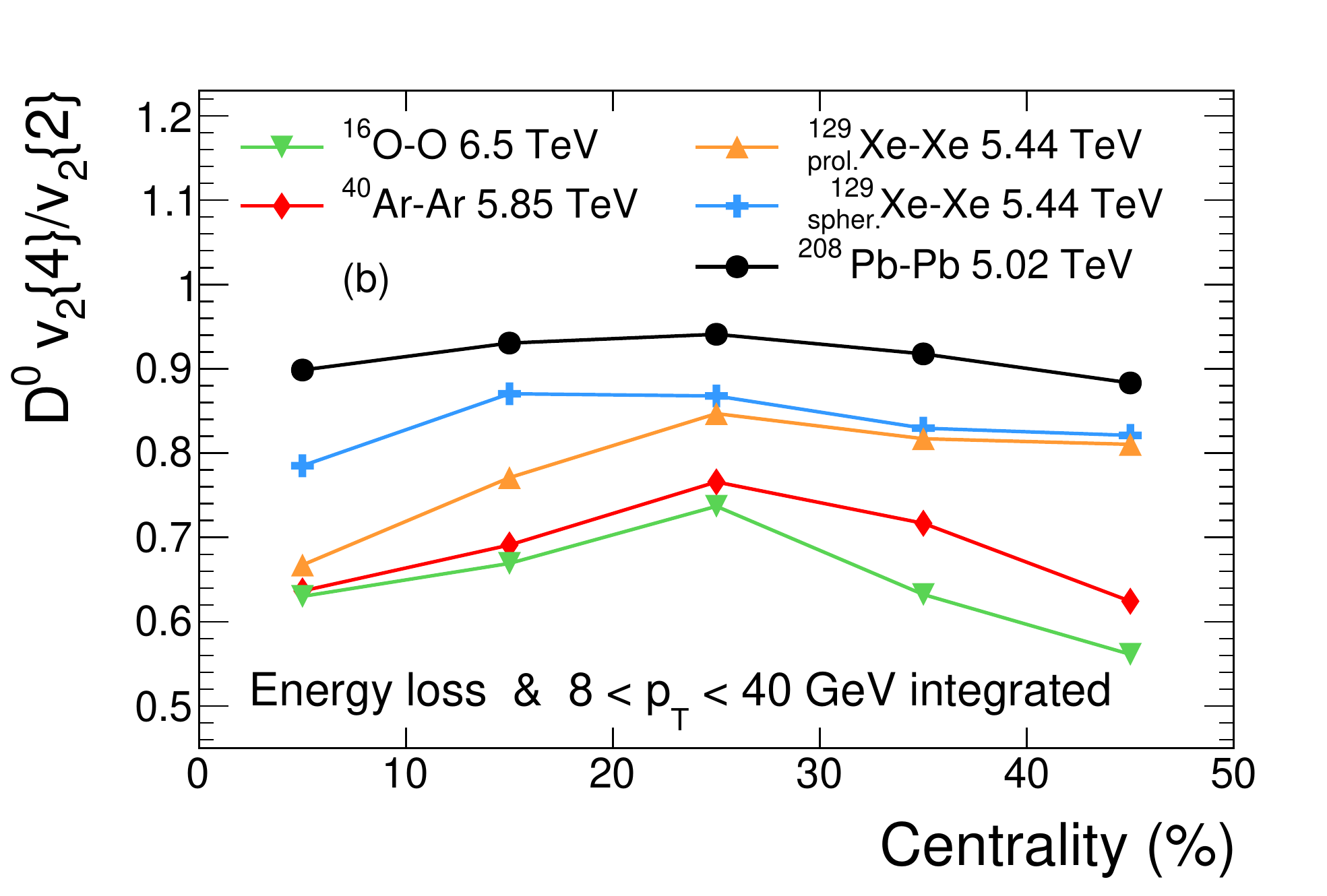}
    \caption{$\Dzero$ meson Langevin low $\pt$ (a) and energy loss high $\pt$ (b) integrated $\vn2\{4\}/\vn2\{2\}$ for $\PbPb$, $\XeXe$ with spherical and prolate initial nuclei, $\ArAr$, and $\OO$ collisions at the \lhc\ top energies.}
    \label{fig:v24v22}
\end{figure}

Finally, in Fig.~\ref{fig:v24v22} we compare the multiparticle cumulants ratio $\vn2\{4\}/\vn2\{2\}$ where for the 4-particle cumulant one correlates one heavy particle and three soft ones.  These were first proposed in~\cite{prado:2016szr,betz:2016ayq} and have yet to be measured in the heavy flavor sector. In~\cite{Katz:2019fkc} it was shown this ratio was mostly dependent on the type of soft initial fluctuations used.  In Fig.~\ref{fig:v24v22} we find that generally $\vn2\{4\}/\vn2\{2\}$ is suppressed with decreasing system size, which is in line with eccentricity calculations from~\cite{Sievert:2019zjr}. Though the lower $\pt$ cut is generally less sensitive to system size, high $\pt$ $\vn2$ fluctuations appear to be significantly more affected by system size (whatever the considered transport model).\\

\begin{figure}[h!]
    \centering
\vspace{3mm}
    \includegraphics[width=0.50\textwidth]{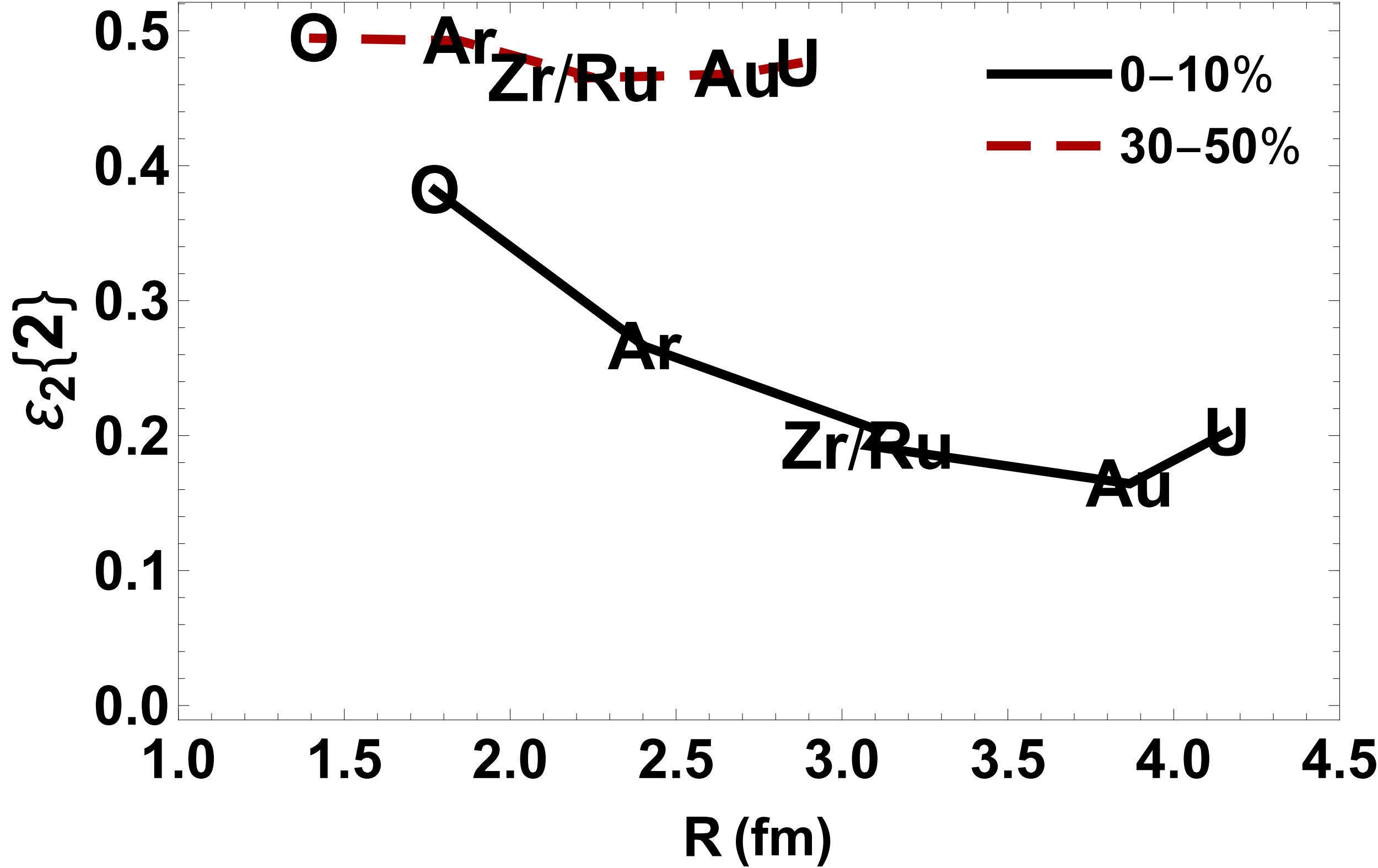}
    \caption{$\varepsilon_2\{2\}$ versus radius for the proposed system size scan at \rhic\ in 0--10\% and 30--50\% centrality classes.}
    \label{fig:eccRHIC}
\end{figure}

There has been a proposal for an intermediate system size scan at \rhic~\cite{Huang:2019tgz} (s\phenix\ would also have these capabilities) and we anticipate a similar effect at these lower energies as well, as shown in Fig.~\ref{fig:eccRHIC}.  To see this directly we also calculate the $\varepsilon_2\{2\}$ vs.~radius relationship at \rhic\ using ions that have been previously ran ($^{238}\U$ and $^{197}\Au$), the isobar run ($^{96}\Ru$ and $^{96}\Zr$), and the proposed intermediate system size scan ($^{40}\Ar$ and $^{16}\Oxigen$).  Generally, we find the same inverse relationship between $\varepsilon_2\{2\}$ and radius in 0--10\% centralities (with the exception of Uranium due to its deformed structure) and the same constant eccentricities at 30--50\% centralities.  Thus, the proposal for a system size scan at \rhic\ should see a similar effect on the $\Dmeson$ mesons as we have shown here but with the added consideration of a lower beam energy.\\

\noindent \textsl{4. Conclusions.} In this letter we make the first predictions for the $\Dmeson$ meson observables for the proposed intermediate system size scan at the \lhc.  We predict that $\raa \rightarrow 1$ gradually as the system size is decreased with mid-central collisions approaching unity before central collisions (as expected due to their smaller system size).  In mid-central collisions we find a clear suppression of $\vnn\{2\}$ in small systems, which shows the significant role played by the system size itself as the geometry of the initial conditions is nearly identical over different systems.  However, in central collisions the eccentricities become larger with a shrinking system size, which cancels out or overcomes the usual suppression effects on $\vn2$.  Thus, we predict that in central collisions $\vn2\{2\}(\pt)$ will be roughly constant across the system size scan.  While the triangular eccentricities in central collisions increase with decreasing system size, triangular azimuthal anisotropy is more sensitive to the system size itself and, thus, one should observe a system size hierarchy. Additionally, we find that $\vn3$ in small systems decreases with increasing centrality, whereas it is known to be roughly constant in $\PbPb$ collisions, which can now be explained by a balance between path length and eccentricity variations with centrality.  Finally, while the average magnitude shifts somewhat between the two, we find that the vast majority of these features are generic for our best fit Langevin and energy loss models.  

If confirmed, these results can help to elucidate the nature of hard probes in small systems. Heavy quarks still lose energy in the system but as the system size path length is decreased they loose significantly less energy.  However, this does not appear to affect $\vn2$ in central collisions because of a significant enhancement of the ellipticity of the initial state. In future work, we hope to explore further soft-heavy correlations system size dependence such as \cite{Plumari:2019yhg}.   As was previously shown in~\cite{Nahrgang:2014vza,Katz:2019fkc,prado:2016szr,Andres:2019eus} both the initial time and decoupling temperature play a significantly role in the $\vnn$'s in the hard sector, thus, indicating that the lifetime of the system matters (especially for $\vn3$).  We point out that these results fall out naturally with an initial conditions+hydrodynamics+heavy flavor Langevin or energy loss scenario and appear to be consistent with $\pPb$ results as well, which would provide further evidence that a Quark-Gluon Plasma could also be formed in smaller systems. However, we have not yet tested this calculation and other studies \cite{Xu:2015iha} suggest that the $\raa$ do not reach unity enough in $\pPb$ to be consistent with data, showing that still remain some open questions.

\noindent \textsl{Acknowledgments.} We wish to thank Z.~Chen, W.~Li, and D.~Perepelitsa for fruitful discussions and help.  The authors thank Funda\c{c}\~ao de Amparo \`a Pesquisa do Estado de S\~ao Paulo (FAPESP) and Conselho Nacional de Desenvolvimento Cient\'ifico e Tecnol\'ogico (CNPq) for support. R.K. is supported by the Region Pays de la Loire (France) under contract No. 2015-08473. C.A.G.P. is supported by the NSFC under grant No. 11521064, MOST of China under Project No. 2014CB845404.  J.N.H. acknowledges the support of the Alfred P. Sloan Foundation, support from the US-DOE Nuclear Science Grant No. DE-SC0019175, and the Office of Advanced Research Computing (OARC) at Rutgers, The State University of New Jersey for providing access to the Amarel cluster and associated research computing resources that have contributed to the results reported here.

\bibliography{BIG}

\begin{thebibliography}{101}%
\makeatletter
\providecommand \@ifxundefined [1]{%
 \@ifx{#1\undefined}
}%
\providecommand \@ifnum [1]{%
 \ifnum #1\expandafter \@firstoftwo
 \else \expandafter \@secondoftwo
 \fi
}%
\providecommand \@ifx [1]{%
 \ifx #1\expandafter \@firstoftwo
 \else \expandafter \@secondoftwo
 \fi
}%
\providecommand \natexlab [1]{#1}%
\providecommand \enquote  [1]{``#1''}%
\providecommand \bibnamefont  [1]{#1}%
\providecommand \bibfnamefont [1]{#1}%
\providecommand \citenamefont [1]{#1}%
\providecommand \href@noop [0]{\@secondoftwo}%
\providecommand \href [0]{\begingroup \@sanitize@url \@href}%
\providecommand \@href[1]{\@@startlink{#1}\@@href}%
\providecommand \@@href[1]{\endgroup#1\@@endlink}%
\providecommand \@sanitize@url [0]{\catcode `\\12\catcode `\$12\catcode
  `\&12\catcode `\#12\catcode `\^12\catcode `\_12\catcode `\%12\relax}%
\providecommand \@@startlink[1]{}%
\providecommand \@@endlink[0]{}%
\providecommand \url  [0]{\begingroup\@sanitize@url \@url }%
\providecommand \@url [1]{\endgroup\@href {#1}{\urlprefix }}%
\providecommand \urlprefix  [0]{URL }%
\providecommand \Eprint [0]{\href }%
\providecommand \doibase [0]{http://dx.doi.org/}%
\providecommand \selectlanguage [0]{\@gobble}%
\providecommand \bibinfo  [0]{\@secondoftwo}%
\providecommand \bibfield  [0]{\@secondoftwo}%
\providecommand \translation [1]{[#1]}%
\providecommand \BibitemOpen [0]{}%
\providecommand \bibitemStop [0]{}%
\providecommand \bibitemNoStop [0]{.\EOS\space}%
\providecommand \EOS [0]{\spacefactor3000\relax}%
\providecommand \BibitemShut  [1]{\csname bibitem#1\endcsname}%
\let\auto@bib@innerbib\@empty
\bibitem [{\citenamefont {Chatrchyan}\ \emph {et~al.}(2013)\citenamefont
  {Chatrchyan} \emph {et~al.}}]{Chatrchyan:2013nka}%
  \BibitemOpen
  \bibfield  {author} {\bibinfo {author} {\bibfnamefont {S.}~\bibnamefont
  {Chatrchyan}} \emph {et~al.} (\bibinfo {collaboration} {CMS}),\ }\href
  {\doibase 10.1016/j.physletb.2013.06.028} {\bibfield  {journal} {\bibinfo
  {journal} {Phys. Lett.}\ }\textbf {\bibinfo {volume} {B724}},\ \bibinfo
  {pages} {213} (\bibinfo {year} {2013})},\ \Eprint
  {http://arxiv.org/abs/1305.0609} {arXiv:1305.0609 [nucl-ex]} \BibitemShut
  {NoStop}%
\bibitem [{\citenamefont {Aaboud}\ \emph {et~al.}(2017)\citenamefont {Aaboud}
  \emph {et~al.}}]{Aaboud:2017acw}%
  \BibitemOpen
  \bibfield  {author} {\bibinfo {author} {\bibfnamefont {M.}~\bibnamefont
  {Aaboud}} \emph {et~al.} (\bibinfo {collaboration} {ATLAS}),\ }\href
  {\doibase 10.1140/epjc/s10052-017-4988-1} {\bibfield  {journal} {\bibinfo
  {journal} {Eur. Phys. J.}\ }\textbf {\bibinfo {volume} {C77}},\ \bibinfo
  {pages} {428} (\bibinfo {year} {2017})},\ \Eprint
  {http://arxiv.org/abs/1705.04176} {arXiv:1705.04176 [hep-ex]} \BibitemShut
  {NoStop}%
\bibitem [{\citenamefont {Aaboud}\ \emph
  {et~al.}(2018{\natexlab{a}})\citenamefont {Aaboud} \emph
  {et~al.}}]{Aaboud:2017blb}%
  \BibitemOpen
  \bibfield  {author} {\bibinfo {author} {\bibfnamefont {M.}~\bibnamefont
  {Aaboud}} \emph {et~al.} (\bibinfo {collaboration} {ATLAS}),\ }\href
  {\doibase 10.1103/PhysRevC.97.024904} {\bibfield  {journal} {\bibinfo
  {journal} {Phys. Rev.}\ }\textbf {\bibinfo {volume} {C97}},\ \bibinfo {pages}
  {024904} (\bibinfo {year} {2018}{\natexlab{a}})},\ \Eprint
  {http://arxiv.org/abs/1708.03559} {arXiv:1708.03559 [hep-ex]} \BibitemShut
  {NoStop}%
\bibitem [{\citenamefont {Aad}\ \emph {et~al.}(2013)\citenamefont {Aad} \emph
  {et~al.}}]{Aad:2013fja}%
  \BibitemOpen
  \bibfield  {author} {\bibinfo {author} {\bibfnamefont {G.}~\bibnamefont
  {Aad}} \emph {et~al.} (\bibinfo {collaboration} {ATLAS}),\ }\href {\doibase
  10.1016/j.physletb.2013.06.057} {\bibfield  {journal} {\bibinfo  {journal}
  {Phys. Lett.}\ }\textbf {\bibinfo {volume} {B725}},\ \bibinfo {pages} {60}
  (\bibinfo {year} {2013})},\ \Eprint {http://arxiv.org/abs/1303.2084}
  {arXiv:1303.2084 [hep-ex]} \BibitemShut {NoStop}%
\bibitem [{\citenamefont {Sirunyan}\ \emph
  {et~al.}(2018{\natexlab{a}})\citenamefont {Sirunyan} \emph
  {et~al.}}]{sirunyan:2018toe}%
  \BibitemOpen
  \bibfield  {author} {\bibinfo {author} {\bibfnamefont {A.~M.}\ \bibnamefont
  {Sirunyan}} \emph {et~al.} (\bibinfo {collaboration} {CMS}),\ }\href
  {\doibase 10.1103/PhysRevLett.121.082301} {\bibfield  {journal} {\bibinfo
  {journal} {Phys. Rev. Lett.}\ }\textbf {\bibinfo {volume} {121}},\ \bibinfo
  {pages} {082301} (\bibinfo {year} {2018}{\natexlab{a}})},\ \Eprint
  {http://arxiv.org/abs/1804.09767} {arXiv:1804.09767 [hep-ex]} \BibitemShut
  {NoStop}%
\bibitem [{\citenamefont {Khachatryan}\ \emph
  {et~al.}(2015{\natexlab{a}})\citenamefont {Khachatryan} \emph
  {et~al.}}]{Khachatryan:2014jra}%
  \BibitemOpen
  \bibfield  {author} {\bibinfo {author} {\bibfnamefont {V.}~\bibnamefont
  {Khachatryan}} \emph {et~al.} (\bibinfo {collaboration} {CMS}),\ }\href
  {\doibase 10.1016/j.physletb.2015.01.034} {\bibfield  {journal} {\bibinfo
  {journal} {Phys. Lett.}\ }\textbf {\bibinfo {volume} {B742}},\ \bibinfo
  {pages} {200} (\bibinfo {year} {2015}{\natexlab{a}})},\ \Eprint
  {http://arxiv.org/abs/1409.3392} {arXiv:1409.3392 [nucl-ex]} \BibitemShut
  {NoStop}%
\bibitem [{\citenamefont {Khachatryan}\ \emph
  {et~al.}(2015{\natexlab{b}})\citenamefont {Khachatryan} \emph
  {et~al.}}]{Khachatryan:2015waa}%
  \BibitemOpen
  \bibfield  {author} {\bibinfo {author} {\bibfnamefont {V.}~\bibnamefont
  {Khachatryan}} \emph {et~al.} (\bibinfo {collaboration} {CMS}),\ }\href
  {\doibase 10.1103/PhysRevLett.115.012301} {\bibfield  {journal} {\bibinfo
  {journal} {Phys. Rev. Lett.}\ }\textbf {\bibinfo {volume} {115}},\ \bibinfo
  {pages} {012301} (\bibinfo {year} {2015}{\natexlab{b}})},\ \Eprint
  {http://arxiv.org/abs/1502.05382} {arXiv:1502.05382 [nucl-ex]} \BibitemShut
  {NoStop}%
\bibitem [{\citenamefont {Khachatryan}\ \emph
  {et~al.}(2015{\natexlab{c}})\citenamefont {Khachatryan} \emph
  {et~al.}}]{Khachatryan:2015oea}%
  \BibitemOpen
  \bibfield  {author} {\bibinfo {author} {\bibfnamefont {V.}~\bibnamefont
  {Khachatryan}} \emph {et~al.} (\bibinfo {collaboration} {CMS}),\ }\href
  {\doibase 10.1103/PhysRevC.92.034911} {\bibfield  {journal} {\bibinfo
  {journal} {Phys. Rev.}\ }\textbf {\bibinfo {volume} {C92}},\ \bibinfo {pages}
  {034911} (\bibinfo {year} {2015}{\natexlab{c}})},\ \Eprint
  {http://arxiv.org/abs/1503.01692} {arXiv:1503.01692 [nucl-ex]} \BibitemShut
  {NoStop}%
\bibitem [{\citenamefont {Sirunyan}\ \emph
  {et~al.}(2018{\natexlab{b}})\citenamefont {Sirunyan} \emph
  {et~al.}}]{Sirunyan:2017uyl}%
  \BibitemOpen
  \bibfield  {author} {\bibinfo {author} {\bibfnamefont {A.~M.}\ \bibnamefont
  {Sirunyan}} \emph {et~al.} (\bibinfo {collaboration} {CMS}),\ }\href
  {\doibase 10.1103/PhysRevLett.120.092301} {\bibfield  {journal} {\bibinfo
  {journal} {Phys. Rev. Lett.}\ }\textbf {\bibinfo {volume} {120}},\ \bibinfo
  {pages} {092301} (\bibinfo {year} {2018}{\natexlab{b}})},\ \Eprint
  {http://arxiv.org/abs/1709.09189} {arXiv:1709.09189 [nucl-ex]} \BibitemShut
  {NoStop}%
\bibitem [{\citenamefont {Abelev}\ \emph
  {et~al.}(2013{\natexlab{a}})\citenamefont {Abelev} \emph
  {et~al.}}]{ABELEV:2013wsa}%
  \BibitemOpen
  \bibfield  {author} {\bibinfo {author} {\bibfnamefont {B.~B.}\ \bibnamefont
  {Abelev}} \emph {et~al.} (\bibinfo {collaboration} {ALICE}),\ }\href
  {\doibase 10.1016/j.physletb.2013.08.024} {\bibfield  {journal} {\bibinfo
  {journal} {Phys. Lett.}\ }\textbf {\bibinfo {volume} {B726}},\ \bibinfo
  {pages} {164} (\bibinfo {year} {2013}{\natexlab{a}})},\ \Eprint
  {http://arxiv.org/abs/1307.3237} {arXiv:1307.3237 [nucl-ex]} \BibitemShut
  {NoStop}%
\bibitem [{\citenamefont {Abelev}\ \emph {et~al.}(2014)\citenamefont {Abelev}
  \emph {et~al.}}]{Abelev:2014mda}%
  \BibitemOpen
  \bibfield  {author} {\bibinfo {author} {\bibfnamefont {B.~B.}\ \bibnamefont
  {Abelev}} \emph {et~al.} (\bibinfo {collaboration} {ALICE}),\ }\href
  {\doibase 10.1103/PhysRevC.90.054901} {\bibfield  {journal} {\bibinfo
  {journal} {Phys. Rev.}\ }\textbf {\bibinfo {volume} {C90}},\ \bibinfo {pages}
  {054901} (\bibinfo {year} {2014})},\ \Eprint {http://arxiv.org/abs/1406.2474}
  {arXiv:1406.2474 [nucl-ex]} \BibitemShut {NoStop}%
\bibitem [{\citenamefont {Adare}\ \emph {et~al.}(2013)\citenamefont {Adare}
  \emph {et~al.}}]{Adare:2013piz}%
  \BibitemOpen
  \bibfield  {author} {\bibinfo {author} {\bibfnamefont {A.}~\bibnamefont
  {Adare}} \emph {et~al.} (\bibinfo {collaboration} {PHENIX}),\ }\href
  {\doibase 10.1103/PhysRevLett.111.212301} {\bibfield  {journal} {\bibinfo
  {journal} {Phys. Rev. Lett.}\ }\textbf {\bibinfo {volume} {111}},\ \bibinfo
  {pages} {212301} (\bibinfo {year} {2013})},\ \Eprint
  {http://arxiv.org/abs/1303.1794} {arXiv:1303.1794 [nucl-ex]} \BibitemShut
  {NoStop}%
\bibitem [{\citenamefont {Adare}\ \emph
  {et~al.}(2015{\natexlab{a}})\citenamefont {Adare} \emph
  {et~al.}}]{Adare:2014keg}%
  \BibitemOpen
  \bibfield  {author} {\bibinfo {author} {\bibfnamefont {A.}~\bibnamefont
  {Adare}} \emph {et~al.} (\bibinfo {collaboration} {PHENIX}),\ }\href
  {\doibase 10.1103/PhysRevLett.114.192301} {\bibfield  {journal} {\bibinfo
  {journal} {Phys. Rev. Lett.}\ }\textbf {\bibinfo {volume} {114}},\ \bibinfo
  {pages} {192301} (\bibinfo {year} {2015}{\natexlab{a}})},\ \Eprint
  {http://arxiv.org/abs/1404.7461} {arXiv:1404.7461 [nucl-ex]} \BibitemShut
  {NoStop}%
\bibitem [{\citenamefont {Aidala}\ \emph
  {et~al.}(2018{\natexlab{a}})\citenamefont {Aidala} \emph
  {et~al.}}]{Aidala:2018mcw}%
  \BibitemOpen
  \bibfield  {author} {\bibinfo {author} {\bibfnamefont {C.}~\bibnamefont
  {Aidala}} \emph {et~al.} (\bibinfo {collaboration} {PHENIX}),\ }\href@noop {}
  {\  (\bibinfo {year} {2018}{\natexlab{a}})},\ \Eprint
  {http://arxiv.org/abs/1805.02973} {arXiv:1805.02973 [nucl-ex]} \BibitemShut
  {NoStop}%
\bibitem [{\citenamefont {Adare}\ \emph
  {et~al.}(2018{\natexlab{a}})\citenamefont {Adare} \emph
  {et~al.}}]{Adare:2018toe}%
  \BibitemOpen
  \bibfield  {author} {\bibinfo {author} {\bibfnamefont {A.}~\bibnamefont
  {Adare}} \emph {et~al.} (\bibinfo {collaboration} {PHENIX}),\ }\href@noop {}
  {\  (\bibinfo {year} {2018}{\natexlab{a}})},\ \Eprint
  {http://arxiv.org/abs/1807.11928} {arXiv:1807.11928 [nucl-ex]} \BibitemShut
  {NoStop}%
\bibitem [{\citenamefont {Adare}\ \emph
  {et~al.}(2015{\natexlab{b}})\citenamefont {Adare} \emph
  {et~al.}}]{Adare:2015ctn}%
  \BibitemOpen
  \bibfield  {author} {\bibinfo {author} {\bibfnamefont {A.}~\bibnamefont
  {Adare}} \emph {et~al.} (\bibinfo {collaboration} {PHENIX}),\ }\href
  {\doibase 10.1103/PhysRevLett.115.142301} {\bibfield  {journal} {\bibinfo
  {journal} {Phys. Rev. Lett.}\ }\textbf {\bibinfo {volume} {115}},\ \bibinfo
  {pages} {142301} (\bibinfo {year} {2015}{\natexlab{b}})},\ \Eprint
  {http://arxiv.org/abs/1507.06273} {arXiv:1507.06273 [nucl-ex]} \BibitemShut
  {NoStop}%
\bibitem [{\citenamefont {Aidala}\ \emph
  {et~al.}(2017{\natexlab{a}})\citenamefont {Aidala} \emph
  {et~al.}}]{Aidala:2016vgl}%
  \BibitemOpen
  \bibfield  {author} {\bibinfo {author} {\bibfnamefont {C.}~\bibnamefont
  {Aidala}} \emph {et~al.},\ }\href {\doibase 10.1103/PhysRevC.95.034910}
  {\bibfield  {journal} {\bibinfo  {journal} {Phys. Rev.}\ }\textbf {\bibinfo
  {volume} {C95}},\ \bibinfo {pages} {034910} (\bibinfo {year}
  {2017}{\natexlab{a}})},\ \Eprint {http://arxiv.org/abs/1609.02894}
  {arXiv:1609.02894 [nucl-ex]} \BibitemShut {NoStop}%
\bibitem [{\citenamefont {Adare}\ \emph
  {et~al.}(2018{\natexlab{b}})\citenamefont {Adare} \emph
  {et~al.}}]{Adare:2017wlc}%
  \BibitemOpen
  \bibfield  {author} {\bibinfo {author} {\bibfnamefont {A.}~\bibnamefont
  {Adare}} \emph {et~al.} (\bibinfo {collaboration} {PHENIX}),\ }\href
  {\doibase 10.1103/PhysRevC.97.064904} {\bibfield  {journal} {\bibinfo
  {journal} {Phys. Rev.}\ }\textbf {\bibinfo {volume} {C97}},\ \bibinfo {pages}
  {064904} (\bibinfo {year} {2018}{\natexlab{b}})},\ \Eprint
  {http://arxiv.org/abs/1710.09736} {arXiv:1710.09736 [nucl-ex]} \BibitemShut
  {NoStop}%
\bibitem [{\citenamefont {Adare}\ \emph
  {et~al.}(2018{\natexlab{c}})\citenamefont {Adare} \emph
  {et~al.}}]{Adare:2017rdq}%
  \BibitemOpen
  \bibfield  {author} {\bibinfo {author} {\bibfnamefont {A.}~\bibnamefont
  {Adare}} \emph {et~al.} (\bibinfo {collaboration} {PHENIX}),\ }\href
  {\doibase 10.1103/PhysRevC.98.014912} {\bibfield  {journal} {\bibinfo
  {journal} {Phys. Rev.}\ }\textbf {\bibinfo {volume} {C98}},\ \bibinfo {pages}
  {014912} (\bibinfo {year} {2018}{\natexlab{c}})},\ \Eprint
  {http://arxiv.org/abs/1711.09003} {arXiv:1711.09003 [hep-ex]} \BibitemShut
  {NoStop}%
\bibitem [{\citenamefont {Aidala}\ \emph
  {et~al.}(2017{\natexlab{b}})\citenamefont {Aidala} \emph
  {et~al.}}]{Aidala:2017pup}%
  \BibitemOpen
  \bibfield  {author} {\bibinfo {author} {\bibfnamefont {C.}~\bibnamefont
  {Aidala}} \emph {et~al.} (\bibinfo {collaboration} {PHENIX}),\ }\href
  {\doibase 10.1103/PhysRevC.96.064905} {\bibfield  {journal} {\bibinfo
  {journal} {Phys. Rev.}\ }\textbf {\bibinfo {volume} {C96}},\ \bibinfo {pages}
  {064905} (\bibinfo {year} {2017}{\natexlab{b}})},\ \Eprint
  {http://arxiv.org/abs/1708.06983} {arXiv:1708.06983 [nucl-ex]} \BibitemShut
  {NoStop}%
\bibitem [{\citenamefont {Aidala}\ \emph
  {et~al.}(2018{\natexlab{b}})\citenamefont {Aidala} \emph
  {et~al.}}]{Aidala:2017ajz}%
  \BibitemOpen
  \bibfield  {author} {\bibinfo {author} {\bibfnamefont {C.}~\bibnamefont
  {Aidala}} \emph {et~al.} (\bibinfo {collaboration} {PHENIX}),\ }\href
  {\doibase 10.1103/PhysRevLett.120.062302} {\bibfield  {journal} {\bibinfo
  {journal} {Phys. Rev. Lett.}\ }\textbf {\bibinfo {volume} {120}},\ \bibinfo
  {pages} {062302} (\bibinfo {year} {2018}{\natexlab{b}})},\ \Eprint
  {http://arxiv.org/abs/1707.06108} {arXiv:1707.06108 [nucl-ex]} \BibitemShut
  {NoStop}%
\bibitem [{\citenamefont {Adam}\ \emph {et~al.}(2017)\citenamefont {Adam} \emph
  {et~al.}}]{ALICE:2017jyt}%
  \BibitemOpen
  \bibfield  {author} {\bibinfo {author} {\bibfnamefont {J.}~\bibnamefont
  {Adam}} \emph {et~al.} (\bibinfo {collaboration} {ALICE}),\ }\href {\doibase
  10.1038/nphys4111} {\bibfield  {journal} {\bibinfo  {journal} {Nature Phys.}\
  }\textbf {\bibinfo {volume} {13}},\ \bibinfo {pages} {535} (\bibinfo {year}
  {2017})},\ \Eprint {http://arxiv.org/abs/1606.07424} {arXiv:1606.07424
  [nucl-ex]} \BibitemShut {NoStop}%
\bibitem [{\citenamefont {Bozek}(2012)}]{Bozek:2011if}%
  \BibitemOpen
  \bibfield  {author} {\bibinfo {author} {\bibfnamefont {P.}~\bibnamefont
  {Bozek}},\ }\href {\doibase 10.1103/PhysRevC.85.014911} {\bibfield  {journal}
  {\bibinfo  {journal} {Phys. Rev.}\ }\textbf {\bibinfo {volume} {C85}},\
  \bibinfo {pages} {014911} (\bibinfo {year} {2012})},\ \Eprint
  {http://arxiv.org/abs/1112.0915} {arXiv:1112.0915 [hep-ph]} \BibitemShut
  {NoStop}%
\bibitem [{\citenamefont {Bozek}\ and\ \citenamefont
  {Broniowski}(2013{\natexlab{a}})}]{Bozek:2012gr}%
  \BibitemOpen
  \bibfield  {author} {\bibinfo {author} {\bibfnamefont {P.}~\bibnamefont
  {Bozek}}\ and\ \bibinfo {author} {\bibfnamefont {W.}~\bibnamefont
  {Broniowski}},\ }\href {\doibase 10.1016/j.physletb.2012.12.051} {\bibfield
  {journal} {\bibinfo  {journal} {Phys. Lett.}\ }\textbf {\bibinfo {volume}
  {B718}},\ \bibinfo {pages} {1557} (\bibinfo {year} {2013}{\natexlab{a}})},\
  \Eprint {http://arxiv.org/abs/1211.0845} {arXiv:1211.0845 [nucl-th]}
  \BibitemShut {NoStop}%
\bibitem [{\citenamefont {Bozek}\ \emph {et~al.}(2013)\citenamefont {Bozek},
  \citenamefont {Broniowski},\ and\ \citenamefont {Torrieri}}]{Bozek:2013ska}%
  \BibitemOpen
  \bibfield  {author} {\bibinfo {author} {\bibfnamefont {P.}~\bibnamefont
  {Bozek}}, \bibinfo {author} {\bibfnamefont {W.}~\bibnamefont {Broniowski}}, \
  and\ \bibinfo {author} {\bibfnamefont {G.}~\bibnamefont {Torrieri}},\ }\href
  {\doibase 10.1103/PhysRevLett.111.172303} {\bibfield  {journal} {\bibinfo
  {journal} {Phys. Rev. Lett.}\ }\textbf {\bibinfo {volume} {111}},\ \bibinfo
  {pages} {172303} (\bibinfo {year} {2013})},\ \Eprint
  {http://arxiv.org/abs/1307.5060} {arXiv:1307.5060 [nucl-th]} \BibitemShut
  {NoStop}%
\bibitem [{\citenamefont {Bozek}\ and\ \citenamefont
  {Broniowski}(2013{\natexlab{b}})}]{Bozek:2013uha}%
  \BibitemOpen
  \bibfield  {author} {\bibinfo {author} {\bibfnamefont {P.}~\bibnamefont
  {Bozek}}\ and\ \bibinfo {author} {\bibfnamefont {W.}~\bibnamefont
  {Broniowski}},\ }\href {\doibase 10.1103/PhysRevC.88.014903} {\bibfield
  {journal} {\bibinfo  {journal} {Phys. Rev.}\ }\textbf {\bibinfo {volume}
  {C88}},\ \bibinfo {pages} {014903} (\bibinfo {year} {2013}{\natexlab{b}})},\
  \Eprint {http://arxiv.org/abs/1304.3044} {arXiv:1304.3044 [nucl-th]}
  \BibitemShut {NoStop}%
\bibitem [{\citenamefont {Kozlov}\ \emph {et~al.}(2014)\citenamefont {Kozlov},
  \citenamefont {Luzum}, \citenamefont {Denicol}, \citenamefont {Jeon},\ and\
  \citenamefont {Gale}}]{Kozlov:2014fqa}%
  \BibitemOpen
  \bibfield  {author} {\bibinfo {author} {\bibfnamefont {I.}~\bibnamefont
  {Kozlov}}, \bibinfo {author} {\bibfnamefont {M.}~\bibnamefont {Luzum}},
  \bibinfo {author} {\bibfnamefont {G.}~\bibnamefont {Denicol}}, \bibinfo
  {author} {\bibfnamefont {S.}~\bibnamefont {Jeon}}, \ and\ \bibinfo {author}
  {\bibfnamefont {C.}~\bibnamefont {Gale}},\ }\href@noop {} {\  (\bibinfo
  {year} {2014})},\ \Eprint {http://arxiv.org/abs/1405.3976} {arXiv:1405.3976
  [nucl-th]} \BibitemShut {NoStop}%
\bibitem [{\citenamefont {Zhou}\ \emph {et~al.}(2015)\citenamefont {Zhou},
  \citenamefont {Zhu}, \citenamefont {Li},\ and\ \citenamefont
  {Song}}]{Zhou:2015iba}%
  \BibitemOpen
  \bibfield  {author} {\bibinfo {author} {\bibfnamefont {Y.}~\bibnamefont
  {Zhou}}, \bibinfo {author} {\bibfnamefont {X.}~\bibnamefont {Zhu}}, \bibinfo
  {author} {\bibfnamefont {P.}~\bibnamefont {Li}}, \ and\ \bibinfo {author}
  {\bibfnamefont {H.}~\bibnamefont {Song}},\ }\href {\doibase
  10.1103/PhysRevC.91.064908} {\bibfield  {journal} {\bibinfo  {journal} {Phys.
  Rev.}\ }\textbf {\bibinfo {volume} {C91}},\ \bibinfo {pages} {064908}
  (\bibinfo {year} {2015})},\ \Eprint {http://arxiv.org/abs/1503.06986}
  {arXiv:1503.06986 [nucl-th]} \BibitemShut {NoStop}%
\bibitem [{\citenamefont {Zhao}\ \emph {et~al.}(2018)\citenamefont {Zhao},
  \citenamefont {Zhou}, \citenamefont {Xu}, \citenamefont {Deng},\ and\
  \citenamefont {Song}}]{Zhao:2017rgg}%
  \BibitemOpen
  \bibfield  {author} {\bibinfo {author} {\bibfnamefont {W.}~\bibnamefont
  {Zhao}}, \bibinfo {author} {\bibfnamefont {Y.}~\bibnamefont {Zhou}}, \bibinfo
  {author} {\bibfnamefont {H.}~\bibnamefont {Xu}}, \bibinfo {author}
  {\bibfnamefont {W.}~\bibnamefont {Deng}}, \ and\ \bibinfo {author}
  {\bibfnamefont {H.}~\bibnamefont {Song}},\ }\href {\doibase
  10.1016/j.physletb.2018.03.022} {\bibfield  {journal} {\bibinfo  {journal}
  {Phys. Lett.}\ }\textbf {\bibinfo {volume} {B780}},\ \bibinfo {pages} {495}
  (\bibinfo {year} {2018})},\ \Eprint {http://arxiv.org/abs/1801.00271}
  {arXiv:1801.00271 [nucl-th]} \BibitemShut {NoStop}%
\bibitem [{\citenamefont {Mäntysaari}\ \emph {et~al.}(2017)\citenamefont
  {Mäntysaari}, \citenamefont {Schenke}, \citenamefont {Shen},\ and\
  \citenamefont {Tribedy}}]{Mantysaari:2017cni}%
  \BibitemOpen
  \bibfield  {author} {\bibinfo {author} {\bibfnamefont {H.}~\bibnamefont
  {Mäntysaari}}, \bibinfo {author} {\bibfnamefont {B.}~\bibnamefont
  {Schenke}}, \bibinfo {author} {\bibfnamefont {C.}~\bibnamefont {Shen}}, \
  and\ \bibinfo {author} {\bibfnamefont {P.}~\bibnamefont {Tribedy}},\ }\href
  {\doibase 10.1016/j.physletb.2017.07.038} {\bibfield  {journal} {\bibinfo
  {journal} {Phys. Lett.}\ }\textbf {\bibinfo {volume} {B772}},\ \bibinfo
  {pages} {681} (\bibinfo {year} {2017})},\ \Eprint
  {http://arxiv.org/abs/1705.03177} {arXiv:1705.03177 [nucl-th]} \BibitemShut
  {NoStop}%
\bibitem [{\citenamefont {Weller}\ and\ \citenamefont
  {Romatschke}(2017)}]{Weller:2017tsr}%
  \BibitemOpen
  \bibfield  {author} {\bibinfo {author} {\bibfnamefont {R.~D.}\ \bibnamefont
  {Weller}}\ and\ \bibinfo {author} {\bibfnamefont {P.}~\bibnamefont
  {Romatschke}},\ }\href {\doibase 10.1016/j.physletb.2017.09.077} {\bibfield
  {journal} {\bibinfo  {journal} {Phys. Lett.}\ }\textbf {\bibinfo {volume}
  {B774}},\ \bibinfo {pages} {351} (\bibinfo {year} {2017})},\ \Eprint
  {http://arxiv.org/abs/1701.07145} {arXiv:1701.07145 [nucl-th]} \BibitemShut
  {NoStop}%
\bibitem [{\citenamefont {Greif}\ \emph {et~al.}(2017)\citenamefont {Greif},
  \citenamefont {Greiner}, \citenamefont {Schenke}, \citenamefont
  {Schlichting},\ and\ \citenamefont {Xu}}]{Greif:2017bnr}%
  \BibitemOpen
  \bibfield  {author} {\bibinfo {author} {\bibfnamefont {M.}~\bibnamefont
  {Greif}}, \bibinfo {author} {\bibfnamefont {C.}~\bibnamefont {Greiner}},
  \bibinfo {author} {\bibfnamefont {B.}~\bibnamefont {Schenke}}, \bibinfo
  {author} {\bibfnamefont {S.}~\bibnamefont {Schlichting}}, \ and\ \bibinfo
  {author} {\bibfnamefont {Z.}~\bibnamefont {Xu}},\ }\href {\doibase
  10.1103/PhysRevD.96.091504} {\bibfield  {journal} {\bibinfo  {journal} {Phys.
  Rev.}\ }\textbf {\bibinfo {volume} {D96}},\ \bibinfo {pages} {091504}
  (\bibinfo {year} {2017})},\ \Eprint {http://arxiv.org/abs/1708.02076}
  {arXiv:1708.02076 [hep-ph]} \BibitemShut {NoStop}%
\bibitem [{\citenamefont {Schenke}\ \emph {et~al.}(2016)\citenamefont
  {Schenke}, \citenamefont {Schlichting}, \citenamefont {Tribedy},\ and\
  \citenamefont {Venugopalan}}]{Schenke:2016lrs}%
  \BibitemOpen
  \bibfield  {author} {\bibinfo {author} {\bibfnamefont {B.}~\bibnamefont
  {Schenke}}, \bibinfo {author} {\bibfnamefont {S.}~\bibnamefont
  {Schlichting}}, \bibinfo {author} {\bibfnamefont {P.}~\bibnamefont
  {Tribedy}}, \ and\ \bibinfo {author} {\bibfnamefont {R.}~\bibnamefont
  {Venugopalan}},\ }\href {\doibase 10.1103/PhysRevLett.117.162301} {\bibfield
  {journal} {\bibinfo  {journal} {Phys. Rev. Lett.}\ }\textbf {\bibinfo
  {volume} {117}},\ \bibinfo {pages} {162301} (\bibinfo {year} {2016})},\
  \Eprint {http://arxiv.org/abs/1607.02496} {arXiv:1607.02496 [hep-ph]}
  \BibitemShut {NoStop}%
\bibitem [{\citenamefont {Mäntysaari}\ and\ \citenamefont
  {Schenke}(2016)}]{Mantysaari:2016ykx}%
  \BibitemOpen
  \bibfield  {author} {\bibinfo {author} {\bibfnamefont {H.}~\bibnamefont
  {Mäntysaari}}\ and\ \bibinfo {author} {\bibfnamefont {B.}~\bibnamefont
  {Schenke}},\ }\href {\doibase 10.1103/PhysRevLett.117.052301} {\bibfield
  {journal} {\bibinfo  {journal} {Phys. Rev. Lett.}\ }\textbf {\bibinfo
  {volume} {117}},\ \bibinfo {pages} {052301} (\bibinfo {year} {2016})},\
  \Eprint {http://arxiv.org/abs/1603.04349} {arXiv:1603.04349 [hep-ph]}
  \BibitemShut {NoStop}%
\bibitem [{\citenamefont {Albacete}\ \emph {et~al.}(2018)\citenamefont
  {Albacete}, \citenamefont {Petersen},\ and\ \citenamefont
  {Soto-Ontoso}}]{Albacete:2017ajt}%
  \BibitemOpen
  \bibfield  {author} {\bibinfo {author} {\bibfnamefont {J.~L.}\ \bibnamefont
  {Albacete}}, \bibinfo {author} {\bibfnamefont {H.}~\bibnamefont {Petersen}},
  \ and\ \bibinfo {author} {\bibfnamefont {A.}~\bibnamefont {Soto-Ontoso}},\
  }\href {\doibase 10.1016/j.physletb.2018.01.011} {\bibfield  {journal}
  {\bibinfo  {journal} {Phys. Lett.}\ }\textbf {\bibinfo {volume} {B778}},\
  \bibinfo {pages} {128} (\bibinfo {year} {2018})},\ \Eprint
  {http://arxiv.org/abs/1707.05592} {arXiv:1707.05592 [hep-ph]} \BibitemShut
  {NoStop}%
\bibitem [{\citenamefont {Bozek}\ and\ \citenamefont
  {Broniowski}(2018)}]{Bozek:2018xzy}%
  \BibitemOpen
  \bibfield  {author} {\bibinfo {author} {\bibfnamefont {P.}~\bibnamefont
  {Bozek}}\ and\ \bibinfo {author} {\bibfnamefont {W.}~\bibnamefont
  {Broniowski}},\ }\href {\doibase 10.1103/PhysRevLett.121.202301} {\bibfield
  {journal} {\bibinfo  {journal} {Phys. Rev. Lett.}\ }\textbf {\bibinfo
  {volume} {121}},\ \bibinfo {pages} {202301} (\bibinfo {year} {2018})},\
  \Eprint {http://arxiv.org/abs/1808.09840} {arXiv:1808.09840 [nucl-th]}
  \BibitemShut {NoStop}%
\bibitem [{\citenamefont {Noronha-Hostler}\ \emph {et~al.}(2019)\citenamefont
  {Noronha-Hostler}, \citenamefont {Paladino}, \citenamefont {Rao},
  \citenamefont {Sievert},\ and\ \citenamefont
  {Wertepny}}]{Noronha-Hostler:2019ytn}%
  \BibitemOpen
  \bibfield  {author} {\bibinfo {author} {\bibfnamefont {J.}~\bibnamefont
  {Noronha-Hostler}}, \bibinfo {author} {\bibfnamefont {N.}~\bibnamefont
  {Paladino}}, \bibinfo {author} {\bibfnamefont {S.}~\bibnamefont {Rao}},
  \bibinfo {author} {\bibfnamefont {M.~D.}\ \bibnamefont {Sievert}}, \ and\
  \bibinfo {author} {\bibfnamefont {D.~E.}\ \bibnamefont {Wertepny}},\
  }\href@noop {} {\  (\bibinfo {year} {2019})},\ \Eprint
  {http://arxiv.org/abs/1905.13323} {arXiv:1905.13323 [hep-ph]} \BibitemShut
  {NoStop}%
\bibitem [{\citenamefont {Citron}\ \emph {et~al.}(2018)\citenamefont {Citron}
  \emph {et~al.}}]{Citron:2018lsq}%
  \BibitemOpen
  \bibfield  {author} {\bibinfo {author} {\bibfnamefont {Z.}~\bibnamefont
  {Citron}} \emph {et~al.}\ }(\bibinfo {year} {2018})\ \Eprint
  {http://arxiv.org/abs/1812.06772} {arXiv:1812.06772 [hep-ph]} \BibitemShut
  {NoStop}%
\bibitem [{\citenamefont {Sievert}\ and\ \citenamefont
  {Noronha-Hostler}(2019)}]{Sievert:2019zjr}%
  \BibitemOpen
  \bibfield  {author} {\bibinfo {author} {\bibfnamefont {M.~D.}\ \bibnamefont
  {Sievert}}\ and\ \bibinfo {author} {\bibfnamefont {J.}~\bibnamefont
  {Noronha-Hostler}},\ }\href@noop {} {\  (\bibinfo {year} {2019})},\ \Eprint
  {http://arxiv.org/abs/1901.01319} {arXiv:1901.01319 [nucl-th]} \BibitemShut
  {NoStop}%
\bibitem [{\citenamefont {Lim}\ \emph {et~al.}(2018)\citenamefont {Lim},
  \citenamefont {Carlson}, \citenamefont {Loizides}, \citenamefont {Lonardoni},
  \citenamefont {Lynn}, \citenamefont {Nagle}, \citenamefont {Orjuela~Koop},\
  and\ \citenamefont {Ouellette}}]{Lim:2018huo}%
  \BibitemOpen
  \bibfield  {author} {\bibinfo {author} {\bibfnamefont {S.~H.}\ \bibnamefont
  {Lim}}, \bibinfo {author} {\bibfnamefont {J.}~\bibnamefont {Carlson}},
  \bibinfo {author} {\bibfnamefont {C.}~\bibnamefont {Loizides}}, \bibinfo
  {author} {\bibfnamefont {D.}~\bibnamefont {Lonardoni}}, \bibinfo {author}
  {\bibfnamefont {J.~E.}\ \bibnamefont {Lynn}}, \bibinfo {author}
  {\bibfnamefont {J.~L.}\ \bibnamefont {Nagle}}, \bibinfo {author}
  {\bibfnamefont {J.~D.}\ \bibnamefont {Orjuela~Koop}}, \ and\ \bibinfo
  {author} {\bibfnamefont {J.}~\bibnamefont {Ouellette}},\ }\href@noop {} {\
  (\bibinfo {year} {2018})},\ \Eprint {http://arxiv.org/abs/1812.08096}
  {arXiv:1812.08096 [nucl-th]} \BibitemShut {NoStop}%
\bibitem [{\citenamefont {Giannini}\ \emph {et~al.}(2019)\citenamefont
  {Giannini}, \citenamefont {Grassi},\ and\ \citenamefont
  {Luzum}}]{Giannini:2019abh}%
  \BibitemOpen
  \bibfield  {author} {\bibinfo {author} {\bibfnamefont {A.~V.}\ \bibnamefont
  {Giannini}}, \bibinfo {author} {\bibfnamefont {F.}~\bibnamefont {Grassi}}, \
  and\ \bibinfo {author} {\bibfnamefont {M.}~\bibnamefont {Luzum}},\
  }\href@noop {} {\  (\bibinfo {year} {2019})},\ \Eprint
  {http://arxiv.org/abs/1904.11488} {arXiv:1904.11488 [nucl-th]} \BibitemShut
  {NoStop}%
\bibitem [{\citenamefont {Huang}\ \emph {et~al.}(2019)\citenamefont {Huang},
  \citenamefont {Chen}, \citenamefont {Jia},\ and\ \citenamefont
  {Li}}]{Huang:2019tgz}%
  \BibitemOpen
  \bibfield  {author} {\bibinfo {author} {\bibfnamefont {S.}~\bibnamefont
  {Huang}}, \bibinfo {author} {\bibfnamefont {Z.}~\bibnamefont {Chen}},
  \bibinfo {author} {\bibfnamefont {J.}~\bibnamefont {Jia}}, \ and\ \bibinfo
  {author} {\bibfnamefont {W.}~\bibnamefont {Li}},\ }\href@noop {} {\
  (\bibinfo {year} {2019})},\ \Eprint {http://arxiv.org/abs/1904.10415}
  {arXiv:1904.10415 [nucl-ex]} \BibitemShut {NoStop}%
\bibitem [{\citenamefont {Adam}\ \emph {et~al.}(2016)\citenamefont {Adam} \emph
  {et~al.}}]{Adam:2015qda}%
  \BibitemOpen
  \bibfield  {author} {\bibinfo {author} {\bibfnamefont {J.}~\bibnamefont
  {Adam}} \emph {et~al.} (\bibinfo {collaboration} {ALICE}),\ }\href {\doibase
  10.1016/j.physletb.2015.12.067} {\bibfield  {journal} {\bibinfo  {journal}
  {Phys. Lett. B}\ }\textbf {\bibinfo {volume} {754}},\ \bibinfo {pages} {81}
  (\bibinfo {year} {2016})},\ \Eprint {http://arxiv.org/abs/1509.07491}
  {arXiv:1509.07491 [nucl-ex]} \BibitemShut {NoStop}%
\bibitem [{\citenamefont {Acharya}\ \emph {et~al.}(2019)\citenamefont {Acharya}
  \emph {et~al.}}]{Acharya:2019mno}%
  \BibitemOpen
  \bibfield  {author} {\bibinfo {author} {\bibfnamefont {S.}~\bibnamefont
  {Acharya}} \emph {et~al.} (\bibinfo {collaboration} {ALICE}),\ }\href
  {\doibase 10.1007/JHEP12(2019)092} {\bibfield  {journal} {\bibinfo  {journal}
  {JHEP}\ }\textbf {\bibinfo {volume} {12}},\ \bibinfo {pages} {092} (\bibinfo
  {year} {2019})},\ \Eprint {http://arxiv.org/abs/1906.03425} {arXiv:1906.03425
  [nucl-ex]} \BibitemShut {NoStop}%
\bibitem [{\citenamefont {Zigic}\ \emph {et~al.}(2019)\citenamefont {Zigic},
  \citenamefont {Salom}, \citenamefont {Auvinen}, \citenamefont {Djordjevic},\
  and\ \citenamefont {Djordjevic}}]{Zigic:2018ovr}%
  \BibitemOpen
  \bibfield  {author} {\bibinfo {author} {\bibfnamefont {D.}~\bibnamefont
  {Zigic}}, \bibinfo {author} {\bibfnamefont {I.}~\bibnamefont {Salom}},
  \bibinfo {author} {\bibfnamefont {J.}~\bibnamefont {Auvinen}}, \bibinfo
  {author} {\bibfnamefont {M.}~\bibnamefont {Djordjevic}}, \ and\ \bibinfo
  {author} {\bibfnamefont {M.}~\bibnamefont {Djordjevic}},\ }\href {\doibase
  10.1016/j.physletb.2019.02.020} {\bibfield  {journal} {\bibinfo  {journal}
  {Phys. Lett.}\ }\textbf {\bibinfo {volume} {B791}},\ \bibinfo {pages} {236}
  (\bibinfo {year} {2019})},\ \Eprint {http://arxiv.org/abs/1805.04786}
  {arXiv:1805.04786 [nucl-th]} \BibitemShut {NoStop}%
\bibitem [{\citenamefont {Shi}\ \emph {et~al.}(2019)\citenamefont {Shi},
  \citenamefont {Liao},\ and\ \citenamefont {Gyulassy}}]{Shi:2018vys}%
  \BibitemOpen
  \bibfield  {author} {\bibinfo {author} {\bibfnamefont {S.}~\bibnamefont
  {Shi}}, \bibinfo {author} {\bibfnamefont {J.}~\bibnamefont {Liao}}, \ and\
  \bibinfo {author} {\bibfnamefont {M.}~\bibnamefont {Gyulassy}},\ }\href
  {\doibase 10.1088/1674-1137/43/4/044101} {\bibfield  {journal} {\bibinfo
  {journal} {Chin. Phys.}\ }\textbf {\bibinfo {volume} {C43}},\ \bibinfo
  {pages} {044101} (\bibinfo {year} {2019})},\ \Eprint
  {http://arxiv.org/abs/1808.05461} {arXiv:1808.05461 [hep-ph]} \BibitemShut
  {NoStop}%
\bibitem [{\citenamefont {Zhang}\ \emph {et~al.}(2019)\citenamefont {Zhang},
  \citenamefont {Marquet}, \citenamefont {Qin}, \citenamefont {Wei},\ and\
  \citenamefont {Xiao}}]{Zhang:2019dth}%
  \BibitemOpen
  \bibfield  {author} {\bibinfo {author} {\bibfnamefont {C.}~\bibnamefont
  {Zhang}}, \bibinfo {author} {\bibfnamefont {C.}~\bibnamefont {Marquet}},
  \bibinfo {author} {\bibfnamefont {G.-Y.}\ \bibnamefont {Qin}}, \bibinfo
  {author} {\bibfnamefont {S.-Y.}\ \bibnamefont {Wei}}, \ and\ \bibinfo
  {author} {\bibfnamefont {B.-W.}\ \bibnamefont {Xiao}},\ }\href {\doibase
  10.1103/PhysRevLett.122.172302} {\bibfield  {journal} {\bibinfo  {journal}
  {Phys. Rev. Lett.}\ }\textbf {\bibinfo {volume} {122}},\ \bibinfo {pages}
  {172302} (\bibinfo {year} {2019})},\ \Eprint
  {http://arxiv.org/abs/1901.10320} {arXiv:1901.10320 [hep-ph]} \BibitemShut
  {NoStop}%
\bibitem [{\citenamefont {Shen}\ \emph {et~al.}(2016)\citenamefont {Shen},
  \citenamefont {Park}, \citenamefont {Paquet}, \citenamefont {Denicol},
  \citenamefont {Jeon},\ and\ \citenamefont {Gale}}]{Shen:2016egw}%
  \BibitemOpen
  \bibfield  {author} {\bibinfo {author} {\bibfnamefont {C.}~\bibnamefont
  {Shen}}, \bibinfo {author} {\bibfnamefont {C.}~\bibnamefont {Park}}, \bibinfo
  {author} {\bibfnamefont {J.-F.}\ \bibnamefont {Paquet}}, \bibinfo {author}
  {\bibfnamefont {G.~S.}\ \bibnamefont {Denicol}}, \bibinfo {author}
  {\bibfnamefont {S.}~\bibnamefont {Jeon}}, \ and\ \bibinfo {author}
  {\bibfnamefont {C.}~\bibnamefont {Gale}},\ }\bibfield  {booktitle} {\emph
  {\bibinfo {booktitle} {{Proceedings, 25th International Conference on
  Ultra-Relativistic Nucleus-Nucleus Collisions (Quark Matter 2015): Kobe,
  Japan, September 27-October 3, 2015}}},\ }\href {\doibase
  10.1016/j.nuclphysa.2016.02.016} {\bibfield  {journal} {\bibinfo  {journal}
  {Nucl. Phys.}\ }\textbf {\bibinfo {volume} {A956}},\ \bibinfo {pages} {741}
  (\bibinfo {year} {2016})},\ \Eprint {http://arxiv.org/abs/1601.03070}
  {arXiv:1601.03070 [hep-ph]} \BibitemShut {NoStop}%
\bibitem [{\citenamefont {Kang}\ \emph {et~al.}(2015)\citenamefont {Kang},
  \citenamefont {Vitev}, \citenamefont {Wang}, \citenamefont {Xing},\ and\
  \citenamefont {Zhang}}]{Kang:2014hha}%
  \BibitemOpen
  \bibfield  {author} {\bibinfo {author} {\bibfnamefont {Z.-B.}\ \bibnamefont
  {Kang}}, \bibinfo {author} {\bibfnamefont {I.}~\bibnamefont {Vitev}},
  \bibinfo {author} {\bibfnamefont {E.}~\bibnamefont {Wang}}, \bibinfo {author}
  {\bibfnamefont {H.}~\bibnamefont {Xing}}, \ and\ \bibinfo {author}
  {\bibfnamefont {C.}~\bibnamefont {Zhang}},\ }\href {\doibase
  10.1016/j.physletb.2014.11.024} {\bibfield  {journal} {\bibinfo  {journal}
  {Phys. Lett.}\ }\textbf {\bibinfo {volume} {B740}},\ \bibinfo {pages} {23}
  (\bibinfo {year} {2015})},\ \Eprint {http://arxiv.org/abs/1409.2494}
  {arXiv:1409.2494 [hep-ph]} \BibitemShut {NoStop}%
\bibitem [{\citenamefont {Xu}\ \emph {et~al.}(2016)\citenamefont {Xu},
  \citenamefont {Cao}, \citenamefont {Qin}, \citenamefont {Ke}, \citenamefont
  {Nahrgang}, \citenamefont {Auvinen},\ and\ \citenamefont
  {Bass}}]{Xu:2015iha}%
  \BibitemOpen
  \bibfield  {author} {\bibinfo {author} {\bibfnamefont {Y.}~\bibnamefont
  {Xu}}, \bibinfo {author} {\bibfnamefont {S.}~\bibnamefont {Cao}}, \bibinfo
  {author} {\bibfnamefont {G.-Y.}\ \bibnamefont {Qin}}, \bibinfo {author}
  {\bibfnamefont {W.}~\bibnamefont {Ke}}, \bibinfo {author} {\bibfnamefont
  {M.}~\bibnamefont {Nahrgang}}, \bibinfo {author} {\bibfnamefont
  {J.}~\bibnamefont {Auvinen}}, \ and\ \bibinfo {author} {\bibfnamefont
  {S.~A.}\ \bibnamefont {Bass}},\ }\bibfield  {booktitle} {\emph {\bibinfo
  {booktitle} {{Proceedings, 7th International Conference on Hard and
  Electromagnetic Probes of High-Energy Nuclear Collisions (Hard Probes 2015):
  Montréal, Québec, Canada, June 29-July 3, 2015}}},\ }\href {\doibase
  10.1016/j.nuclphysbps.2016.05.050} {\bibfield  {journal} {\bibinfo  {journal}
  {Nucl. Part. Phys. Proc.}\ }\textbf {\bibinfo {volume} {276-278}},\ \bibinfo
  {pages} {225} (\bibinfo {year} {2016})},\ \Eprint
  {http://arxiv.org/abs/1510.07520} {arXiv:1510.07520 [nucl-th]} \BibitemShut
  {NoStop}%
\bibitem [{\citenamefont {Sharma}\ \emph {et~al.}(2009)\citenamefont {Sharma},
  \citenamefont {Vitev},\ and\ \citenamefont {Zhang}}]{Sharma:2009hn}%
  \BibitemOpen
  \bibfield  {author} {\bibinfo {author} {\bibfnamefont {R.}~\bibnamefont
  {Sharma}}, \bibinfo {author} {\bibfnamefont {I.}~\bibnamefont {Vitev}}, \
  and\ \bibinfo {author} {\bibfnamefont {B.-W.}\ \bibnamefont {Zhang}},\ }\href
  {\doibase 10.1103/PhysRevC.80.054902} {\bibfield  {journal} {\bibinfo
  {journal} {Phys. Rev.}\ }\textbf {\bibinfo {volume} {C80}},\ \bibinfo {pages}
  {054902} (\bibinfo {year} {2009})},\ \Eprint {http://arxiv.org/abs/0904.0032}
  {arXiv:0904.0032 [hep-ph]} \BibitemShut {NoStop}%
\bibitem [{\citenamefont {Sirunyan}\ \emph {et~al.}(2017)\citenamefont
  {Sirunyan} \emph {et~al.}}]{sirunyan:2017plt}%
  \BibitemOpen
  \bibfield  {author} {\bibinfo {author} {\bibfnamefont {A.~M.}\ \bibnamefont
  {Sirunyan}} \emph {et~al.} (\bibinfo {collaboration} {CMS}),\ }\href@noop {}
  {\  (\bibinfo {year} {2017})},\ \Eprint {http://arxiv.org/abs/1708.03497}
  {arXiv:1708.03497 [nucl-ex]} \BibitemShut {NoStop}%
\bibitem [{\citenamefont {collaboration}(2017{\natexlab{a}})}]{ATLAS:2017zby}%
  \BibitemOpen
  \bibfield  {author} {\bibinfo {author} {\bibfnamefont {T.~A.}\ \bibnamefont
  {collaboration}} (\bibinfo {collaboration} {ATLAS}),\ }\href@noop {} {\
  (\bibinfo {year} {2017}{\natexlab{a}})}\BibitemShut {NoStop}%
\bibitem [{\citenamefont {Djordjevic}\ \emph {et~al.}(2019)\citenamefont
  {Djordjevic}, \citenamefont {Stojku}, \citenamefont {Djordjevic},\ and\
  \citenamefont {Huovinen}}]{Djordjevic:2019tdu}%
  \BibitemOpen
  \bibfield  {author} {\bibinfo {author} {\bibfnamefont {M.}~\bibnamefont
  {Djordjevic}}, \bibinfo {author} {\bibfnamefont {S.}~\bibnamefont {Stojku}},
  \bibinfo {author} {\bibfnamefont {M.}~\bibnamefont {Djordjevic}}, \ and\
  \bibinfo {author} {\bibfnamefont {P.}~\bibnamefont {Huovinen}},\ }\href@noop
  {} {\  (\bibinfo {year} {2019})},\ \Eprint {http://arxiv.org/abs/1903.06829}
  {arXiv:1903.06829 [hep-ph]} \BibitemShut {NoStop}%
\bibitem [{\citenamefont {Xu}\ \emph {et~al.}(2019)\citenamefont {Xu} \emph
  {et~al.}}]{Xu:2018gux}%
  \BibitemOpen
  \bibfield  {author} {\bibinfo {author} {\bibfnamefont {Y.}~\bibnamefont {Xu}}
  \emph {et~al.},\ }\href {\doibase 10.1103/PhysRevC.99.014902} {\bibfield
  {journal} {\bibinfo  {journal} {Phys. Rev.}\ }\textbf {\bibinfo {volume}
  {C99}},\ \bibinfo {pages} {014902} (\bibinfo {year} {2019})},\ \Eprint
  {http://arxiv.org/abs/1809.10734} {arXiv:1809.10734 [nucl-th]} \BibitemShut
  {NoStop}%
\bibitem [{\citenamefont {Moreland}\ \emph {et~al.}(2015)\citenamefont
  {Moreland}, \citenamefont {Bernhard},\ and\ \citenamefont
  {Bass}}]{Moreland:2014oya}%
  \BibitemOpen
  \bibfield  {author} {\bibinfo {author} {\bibfnamefont {J.~S.}\ \bibnamefont
  {Moreland}}, \bibinfo {author} {\bibfnamefont {J.~E.}\ \bibnamefont
  {Bernhard}}, \ and\ \bibinfo {author} {\bibfnamefont {S.~A.}\ \bibnamefont
  {Bass}},\ }\href {\doibase 10.1103/PhysRevC.92.011901} {\bibfield  {journal}
  {\bibinfo  {journal} {Phys. Rev.}\ }\textbf {\bibinfo {volume} {C92}},\
  \bibinfo {pages} {011901} (\bibinfo {year} {2015})},\ \Eprint
  {http://arxiv.org/abs/1412.4708} {arXiv:1412.4708 [nucl-th]} \BibitemShut
  {NoStop}%
\bibitem [{\citenamefont {Noronha-Hostler}\ \emph {et~al.}(2014)\citenamefont
  {Noronha-Hostler}, \citenamefont {Noronha},\ and\ \citenamefont
  {Grassi}}]{Noronha-Hostler:2014dqa}%
  \BibitemOpen
  \bibfield  {author} {\bibinfo {author} {\bibfnamefont {J.}~\bibnamefont
  {Noronha-Hostler}}, \bibinfo {author} {\bibfnamefont {J.}~\bibnamefont
  {Noronha}}, \ and\ \bibinfo {author} {\bibfnamefont {F.}~\bibnamefont
  {Grassi}},\ }\href {\doibase 10.1103/PhysRevC.90.034907} {\bibfield
  {journal} {\bibinfo  {journal} {Phys. Rev.}\ }\textbf {\bibinfo {volume}
  {C90}},\ \bibinfo {pages} {034907} (\bibinfo {year} {2014})},\ \Eprint
  {http://arxiv.org/abs/1406.3333} {arXiv:1406.3333 [nucl-th]} \BibitemShut
  {NoStop}%
\bibitem [{\citenamefont {Noronha-Hostler}\ \emph {et~al.}(2013)\citenamefont
  {Noronha-Hostler}, \citenamefont {Denicol}, \citenamefont {Noronha},
  \citenamefont {Andrade},\ and\ \citenamefont
  {Grassi}}]{Noronha-Hostler:2013gga}%
  \BibitemOpen
  \bibfield  {author} {\bibinfo {author} {\bibfnamefont {J.}~\bibnamefont
  {Noronha-Hostler}}, \bibinfo {author} {\bibfnamefont {G.~S.}\ \bibnamefont
  {Denicol}}, \bibinfo {author} {\bibfnamefont {J.}~\bibnamefont {Noronha}},
  \bibinfo {author} {\bibfnamefont {R.~P.~G.}\ \bibnamefont {Andrade}}, \ and\
  \bibinfo {author} {\bibfnamefont {F.}~\bibnamefont {Grassi}},\ }\href
  {\doibase 10.1103/PhysRevC.88.044916} {\bibfield  {journal} {\bibinfo
  {journal} {Phys. Rev.}\ }\textbf {\bibinfo {volume} {C88}},\ \bibinfo {pages}
  {044916} (\bibinfo {year} {2013})},\ \Eprint {http://arxiv.org/abs/1305.1981}
  {arXiv:1305.1981 [nucl-th]} \BibitemShut {NoStop}%
\bibitem [{\citenamefont {Prado}\ \emph {et~al.}(2017)\citenamefont {Prado},
  \citenamefont {Noronha-Hostler}, \citenamefont {Katz}, \citenamefont
  {Suaide}, \citenamefont {Noronha}, \citenamefont {Munhoz},\ and\
  \citenamefont {Cosentino}}]{prado:2016szr}%
  \BibitemOpen
  \bibfield  {author} {\bibinfo {author} {\bibfnamefont {C.~A.~G.}\
  \bibnamefont {Prado}}, \bibinfo {author} {\bibfnamefont {J.}~\bibnamefont
  {Noronha-Hostler}}, \bibinfo {author} {\bibfnamefont {R.}~\bibnamefont
  {Katz}}, \bibinfo {author} {\bibfnamefont {A.~A.~P.}\ \bibnamefont {Suaide}},
  \bibinfo {author} {\bibfnamefont {J.}~\bibnamefont {Noronha}}, \bibinfo
  {author} {\bibfnamefont {M.~G.}\ \bibnamefont {Munhoz}}, \ and\ \bibinfo
  {author} {\bibfnamefont {M.~R.}\ \bibnamefont {Cosentino}},\ }\href {\doibase
  10.1103/PhysRevC.96.064903} {\bibfield  {journal} {\bibinfo  {journal} {Phys.
  Rev.}\ }\textbf {\bibinfo {volume} {C96}},\ \bibinfo {pages} {064903}
  (\bibinfo {year} {2017})},\ \Eprint {http://arxiv.org/abs/1611.02965}
  {arXiv:1611.02965 [nucl-th]} \BibitemShut {NoStop}%
\bibitem [{\citenamefont {Alba}\ \emph {et~al.}(2018)\citenamefont {Alba},
  \citenamefont {Mantovani~Sarti}, \citenamefont {Noronha}, \citenamefont
  {Noronha-Hostler}, \citenamefont {Parotto}, \citenamefont
  {Portillo~Vazquez},\ and\ \citenamefont {Ratti}}]{Alba:2017hhe}%
  \BibitemOpen
  \bibfield  {author} {\bibinfo {author} {\bibfnamefont {P.}~\bibnamefont
  {Alba}}, \bibinfo {author} {\bibfnamefont {V.}~\bibnamefont
  {Mantovani~Sarti}}, \bibinfo {author} {\bibfnamefont {J.}~\bibnamefont
  {Noronha}}, \bibinfo {author} {\bibfnamefont {J.}~\bibnamefont
  {Noronha-Hostler}}, \bibinfo {author} {\bibfnamefont {P.}~\bibnamefont
  {Parotto}}, \bibinfo {author} {\bibfnamefont {I.}~\bibnamefont
  {Portillo~Vazquez}}, \ and\ \bibinfo {author} {\bibfnamefont
  {C.}~\bibnamefont {Ratti}},\ }\href {\doibase 10.1103/PhysRevC.98.034909}
  {\bibfield  {journal} {\bibinfo  {journal} {Phys. Rev.}\ }\textbf {\bibinfo
  {volume} {C98}},\ \bibinfo {pages} {034909} (\bibinfo {year} {2018})},\
  \Eprint {http://arxiv.org/abs/1711.05207} {arXiv:1711.05207 [nucl-th]}
  \BibitemShut {NoStop}%
\bibitem [{\citenamefont {Giacalone}\ \emph {et~al.}(2018)\citenamefont
  {Giacalone}, \citenamefont {Noronha-Hostler}, \citenamefont {Luzum},\ and\
  \citenamefont {Ollitrault}}]{Giacalone:2017dud}%
  \BibitemOpen
  \bibfield  {author} {\bibinfo {author} {\bibfnamefont {G.}~\bibnamefont
  {Giacalone}}, \bibinfo {author} {\bibfnamefont {J.}~\bibnamefont
  {Noronha-Hostler}}, \bibinfo {author} {\bibfnamefont {M.}~\bibnamefont
  {Luzum}}, \ and\ \bibinfo {author} {\bibfnamefont {J.-Y.}\ \bibnamefont
  {Ollitrault}},\ }\href {\doibase 10.1103/PhysRevC.97.034904} {\bibfield
  {journal} {\bibinfo  {journal} {Phys. Rev.}\ }\textbf {\bibinfo {volume}
  {C97}},\ \bibinfo {pages} {034904} (\bibinfo {year} {2018})},\ \Eprint
  {http://arxiv.org/abs/1711.08499} {arXiv:1711.08499 [nucl-th]} \BibitemShut
  {NoStop}%
\bibitem [{\citenamefont {Katz}\ \emph {et~al.}(2019)\citenamefont {Katz},
  \citenamefont {Prado}, \citenamefont {Noronha-Hostler}, \citenamefont
  {Noronha},\ and\ \citenamefont {Suaide}}]{Katz:2019fkc}%
  \BibitemOpen
  \bibfield  {author} {\bibinfo {author} {\bibfnamefont {R.}~\bibnamefont
  {Katz}}, \bibinfo {author} {\bibfnamefont {C.~A.~G.}\ \bibnamefont {Prado}},
  \bibinfo {author} {\bibfnamefont {J.}~\bibnamefont {Noronha-Hostler}},
  \bibinfo {author} {\bibfnamefont {J.}~\bibnamefont {Noronha}}, \ and\
  \bibinfo {author} {\bibfnamefont {A.~A.~P.}\ \bibnamefont {Suaide}},\
  }\href@noop {} {\  (\bibinfo {year} {2019})},\ \Eprint
  {http://arxiv.org/abs/1906.10768} {arXiv:1906.10768 [nucl-th]} \BibitemShut
  {NoStop}%
\bibitem [{\citenamefont {Cacciari}\ \emph {et~al.}(1998)\citenamefont
  {Cacciari}, \citenamefont {Greco},\ and\ \citenamefont
  {Nason}}]{Cacciari:1998it}%
  \BibitemOpen
  \bibfield  {author} {\bibinfo {author} {\bibfnamefont {M.}~\bibnamefont
  {Cacciari}}, \bibinfo {author} {\bibfnamefont {M.}~\bibnamefont {Greco}}, \
  and\ \bibinfo {author} {\bibfnamefont {P.}~\bibnamefont {Nason}},\ }\href
  {\doibase 10.1088/1126-6708/1998/05/007} {\bibfield  {journal} {\bibinfo
  {journal} {JHEP}\ }\textbf {\bibinfo {volume} {05}},\ \bibinfo {pages} {007}
  (\bibinfo {year} {1998})},\ \Eprint {http://arxiv.org/abs/hep-ph/9803400}
  {arXiv:hep-ph/9803400 [hep-ph]} \BibitemShut {NoStop}%
\bibitem [{\citenamefont {Cacciari}\ \emph {et~al.}(2001)\citenamefont
  {Cacciari}, \citenamefont {Frixione},\ and\ \citenamefont
  {Nason}}]{Cacciari:2001td}%
  \BibitemOpen
  \bibfield  {author} {\bibinfo {author} {\bibfnamefont {M.}~\bibnamefont
  {Cacciari}}, \bibinfo {author} {\bibfnamefont {S.}~\bibnamefont {Frixione}},
  \ and\ \bibinfo {author} {\bibfnamefont {P.}~\bibnamefont {Nason}},\ }\href
  {\doibase 10.1088/1126-6708/2001/03/006} {\bibfield  {journal} {\bibinfo
  {journal} {JHEP}\ }\textbf {\bibinfo {volume} {03}},\ \bibinfo {pages} {006}
  (\bibinfo {year} {2001})},\ \Eprint {http://arxiv.org/abs/hep-ph/0102134}
  {arXiv:hep-ph/0102134 [hep-ph]} \BibitemShut {NoStop}%
\bibitem [{\citenamefont {Moore}\ and\ \citenamefont
  {Teaney}(2005)}]{Moore:2004tg}%
  \BibitemOpen
  \bibfield  {author} {\bibinfo {author} {\bibfnamefont {G.~D.}\ \bibnamefont
  {Moore}}\ and\ \bibinfo {author} {\bibfnamefont {D.}~\bibnamefont {Teaney}},\
  }\href {\doibase 10.1103/PhysRevC.71.064904} {\bibfield  {journal} {\bibinfo
  {journal} {Phys. Rev.}\ }\textbf {\bibinfo {volume} {C71}},\ \bibinfo {pages}
  {064904} (\bibinfo {year} {2005})},\ \Eprint
  {http://arxiv.org/abs/hep-ph/0412346} {arXiv:hep-ph/0412346 [hep-ph]}
  \BibitemShut {NoStop}%
\bibitem [{\citenamefont {Aaboud}\ \emph
  {et~al.}(2018{\natexlab{b}})\citenamefont {Aaboud} \emph
  {et~al.}}]{Aaboud:2018bdg}%
  \BibitemOpen
  \bibfield  {author} {\bibinfo {author} {\bibfnamefont {M.}~\bibnamefont
  {Aaboud}} \emph {et~al.} (\bibinfo {collaboration} {ATLAS}),\ }\href
  {\doibase 10.1103/PhysRevC.98.044905} {\bibfield  {journal} {\bibinfo
  {journal} {Phys. Rev.}\ }\textbf {\bibinfo {volume} {C98}},\ \bibinfo {pages}
  {044905} (\bibinfo {year} {2018}{\natexlab{b}})},\ \Eprint
  {http://arxiv.org/abs/1805.05220} {arXiv:1805.05220 [nucl-ex]} \BibitemShut
  {NoStop}%
\bibitem [{\citenamefont {Bernhard}\ \emph {et~al.}(2016)\citenamefont
  {Bernhard}, \citenamefont {Moreland}, \citenamefont {Bass}, \citenamefont
  {Liu},\ and\ \citenamefont {Heinz}}]{bernhard:2016tnd}%
  \BibitemOpen
  \bibfield  {author} {\bibinfo {author} {\bibfnamefont {J.~E.}\ \bibnamefont
  {Bernhard}}, \bibinfo {author} {\bibfnamefont {J.~S.}\ \bibnamefont
  {Moreland}}, \bibinfo {author} {\bibfnamefont {S.~A.}\ \bibnamefont {Bass}},
  \bibinfo {author} {\bibfnamefont {J.}~\bibnamefont {Liu}}, \ and\ \bibinfo
  {author} {\bibfnamefont {U.}~\bibnamefont {Heinz}},\ }\href {\doibase
  10.1103/PhysRevC.94.024907} {\bibfield  {journal} {\bibinfo  {journal} {Phys.
  Rev.}\ }\textbf {\bibinfo {volume} {C94}},\ \bibinfo {pages} {024907}
  (\bibinfo {year} {2016})},\ \Eprint {http://arxiv.org/abs/1605.03954}
  {arXiv:1605.03954 [nucl-th]} \BibitemShut {NoStop}%
\bibitem [{\citenamefont {Moller}\ \emph {et~al.}(2016)\citenamefont {Moller},
  \citenamefont {Sierk}, \citenamefont {Ichikawa},\ and\ \citenamefont
  {Sagawa}}]{Moller:2015fba}%
  \BibitemOpen
  \bibfield  {author} {\bibinfo {author} {\bibfnamefont {P.}~\bibnamefont
  {Moller}}, \bibinfo {author} {\bibfnamefont {A.~J.}\ \bibnamefont {Sierk}},
  \bibinfo {author} {\bibfnamefont {T.}~\bibnamefont {Ichikawa}}, \ and\
  \bibinfo {author} {\bibfnamefont {H.}~\bibnamefont {Sagawa}},\ }\href
  {\doibase 10.1016/j.adt.2015.10.002} {\bibfield  {journal} {\bibinfo
  {journal} {Atom. Data Nucl. Data Tabl.}\ }\textbf {\bibinfo {volume}
  {109-110}},\ \bibinfo {pages} {1} (\bibinfo {year} {2016})},\ \Eprint
  {http://arxiv.org/abs/1508.06294} {arXiv:1508.06294 [nucl-th]} \BibitemShut
  {NoStop}%
\bibitem [{\citenamefont {Betz}\ \emph {et~al.}(2017)\citenamefont {Betz},
  \citenamefont {Gyulassy}, \citenamefont {Luzum}, \citenamefont {Noronha},
  \citenamefont {Noronha-Hostler}, \citenamefont {Portillo},\ and\
  \citenamefont {Ratti}}]{betz:2016ayq}%
  \BibitemOpen
  \bibfield  {author} {\bibinfo {author} {\bibfnamefont {B.}~\bibnamefont
  {Betz}}, \bibinfo {author} {\bibfnamefont {M.}~\bibnamefont {Gyulassy}},
  \bibinfo {author} {\bibfnamefont {M.}~\bibnamefont {Luzum}}, \bibinfo
  {author} {\bibfnamefont {J.}~\bibnamefont {Noronha}}, \bibinfo {author}
  {\bibfnamefont {J.}~\bibnamefont {Noronha-Hostler}}, \bibinfo {author}
  {\bibfnamefont {I.}~\bibnamefont {Portillo}}, \ and\ \bibinfo {author}
  {\bibfnamefont {C.}~\bibnamefont {Ratti}},\ }\href {\doibase
  10.1103/PhysRevC.95.044901} {\bibfield  {journal} {\bibinfo  {journal} {Phys.
  Rev.}\ }\textbf {\bibinfo {volume} {C95}},\ \bibinfo {pages} {044901}
  (\bibinfo {year} {2017})},\ \Eprint {http://arxiv.org/abs/1609.05171}
  {arXiv:1609.05171 [nucl-th]} \BibitemShut {NoStop}%
\bibitem [{\citenamefont {Noronha-Hostler}\ \emph {et~al.}(2016)\citenamefont
  {Noronha-Hostler}, \citenamefont {Betz}, \citenamefont {Noronha},\ and\
  \citenamefont {Gyulassy}}]{noronha-hostler:2016eow}%
  \BibitemOpen
  \bibfield  {author} {\bibinfo {author} {\bibfnamefont {J.}~\bibnamefont
  {Noronha-Hostler}}, \bibinfo {author} {\bibfnamefont {B.}~\bibnamefont
  {Betz}}, \bibinfo {author} {\bibfnamefont {J.}~\bibnamefont {Noronha}}, \
  and\ \bibinfo {author} {\bibfnamefont {M.}~\bibnamefont {Gyulassy}},\ }\href
  {\doibase 10.1103/PhysRevLett.116.252301} {\bibfield  {journal} {\bibinfo
  {journal} {Phys. Rev. Lett.}\ }\textbf {\bibinfo {volume} {116}},\ \bibinfo
  {pages} {252301} (\bibinfo {year} {2016})},\ \Eprint
  {http://arxiv.org/abs/1602.03788} {arXiv:1602.03788 [nucl-th]} \BibitemShut
  {NoStop}%
\bibitem [{\citenamefont {Sirunyan}\ \emph
  {et~al.}(2018{\natexlab{c}})\citenamefont {Sirunyan} \emph
  {et~al.}}]{sirunyan:2017pan}%
  \BibitemOpen
  \bibfield  {author} {\bibinfo {author} {\bibfnamefont {A.~M.}\ \bibnamefont
  {Sirunyan}} \emph {et~al.} (\bibinfo {collaboration} {CMS}),\ }\href
  {\doibase 10.1016/j.physletb.2017.11.041} {\bibfield  {journal} {\bibinfo
  {journal} {Phys. Lett.}\ }\textbf {\bibinfo {volume} {B776}},\ \bibinfo
  {pages} {195} (\bibinfo {year} {2018}{\natexlab{c}})},\ \Eprint
  {http://arxiv.org/abs/1702.00630} {arXiv:1702.00630 [hep-ex]} \BibitemShut
  {NoStop}%
\bibitem [{\citenamefont {Andres}\ \emph {et~al.}(2019)\citenamefont {Andres},
  \citenamefont {Armesto}, \citenamefont {Niemi}, \citenamefont {Paatelainen},\
  and\ \citenamefont {Salgado}}]{Andres:2019eus}%
  \BibitemOpen
  \bibfield  {author} {\bibinfo {author} {\bibfnamefont {C.}~\bibnamefont
  {Andres}}, \bibinfo {author} {\bibfnamefont {N.}~\bibnamefont {Armesto}},
  \bibinfo {author} {\bibfnamefont {H.}~\bibnamefont {Niemi}}, \bibinfo
  {author} {\bibfnamefont {R.}~\bibnamefont {Paatelainen}}, \ and\ \bibinfo
  {author} {\bibfnamefont {C.~A.}\ \bibnamefont {Salgado}},\ }\href@noop {} {\
  (\bibinfo {year} {2019})},\ \Eprint {http://arxiv.org/abs/1902.03231}
  {arXiv:1902.03231 [hep-ph]} \BibitemShut {NoStop}%
\bibitem [{\citenamefont {Abelev}\ \emph
  {et~al.}(2013{\natexlab{b}})\citenamefont {Abelev} \emph
  {et~al.}}]{ALICE:2012mj}%
  \BibitemOpen
  \bibfield  {author} {\bibinfo {author} {\bibfnamefont {B.}~\bibnamefont
  {Abelev}} \emph {et~al.} (\bibinfo {collaboration} {ALICE}),\ }\href
  {\doibase 10.1103/PhysRevLett.110.082302} {\bibfield  {journal} {\bibinfo
  {journal} {Phys. Rev. Lett.}\ }\textbf {\bibinfo {volume} {110}},\ \bibinfo
  {pages} {082302} (\bibinfo {year} {2013}{\natexlab{b}})},\ \Eprint
  {http://arxiv.org/abs/1210.4520} {arXiv:1210.4520 [nucl-ex]} \BibitemShut
  {NoStop}%
\bibitem [{\citenamefont {Collaboration}(2013)}]{CMS:2013cka}%
  \BibitemOpen
  \bibfield  {author} {\bibinfo {author} {\bibfnamefont {C.}~\bibnamefont
  {Collaboration}} (\bibinfo {collaboration} {CMS}),\ }\href@noop {} {\
  (\bibinfo {year} {2013})}\BibitemShut {NoStop}%
\bibitem [{\citenamefont {Khachatryan}\ \emph {et~al.}(2017)\citenamefont
  {Khachatryan} \emph {et~al.}}]{Khachatryan:2016odn}%
  \BibitemOpen
  \bibfield  {author} {\bibinfo {author} {\bibfnamefont {V.}~\bibnamefont
  {Khachatryan}} \emph {et~al.} (\bibinfo {collaboration} {CMS}),\ }\href
  {\doibase 10.1007/JHEP04(2017)039} {\bibfield  {journal} {\bibinfo  {journal}
  {JHEP}\ }\textbf {\bibinfo {volume} {04}},\ \bibinfo {pages} {039} (\bibinfo
  {year} {2017})},\ \Eprint {http://arxiv.org/abs/1611.01664} {arXiv:1611.01664
  [nucl-ex]} \BibitemShut {NoStop}%
\bibitem [{\citenamefont {collaboration}(2017{\natexlab{b}})}]{ATLAS:2017dgr}%
  \BibitemOpen
  \bibfield  {author} {\bibinfo {author} {\bibfnamefont {T.~A.}\ \bibnamefont
  {collaboration}} (\bibinfo {collaboration} {ATLAS}),\ }\href@noop {} {\
  (\bibinfo {year} {2017}{\natexlab{b}})}\BibitemShut {NoStop}%
\bibitem [{\citenamefont {Aaboud}\ \emph
  {et~al.}(2018{\natexlab{c}})\citenamefont {Aaboud} \emph
  {et~al.}}]{Aaboud:2017cif}%
  \BibitemOpen
  \bibfield  {author} {\bibinfo {author} {\bibfnamefont {M.}~\bibnamefont
  {Aaboud}} \emph {et~al.} (\bibinfo {collaboration} {ATLAS}),\ }\href
  {\doibase 10.1140/epjc/s10052-018-5624-4} {\bibfield  {journal} {\bibinfo
  {journal} {Eur. Phys. J.}\ }\textbf {\bibinfo {volume} {C78}},\ \bibinfo
  {pages} {171} (\bibinfo {year} {2018}{\natexlab{c}})},\ \Eprint
  {http://arxiv.org/abs/1709.03089} {arXiv:1709.03089 [nucl-ex]} \BibitemShut
  {NoStop}%
\bibitem [{\citenamefont {Bencedi}(2016)}]{Bencedi:2016tks}%
  \BibitemOpen
  \bibfield  {author} {\bibinfo {author} {\bibfnamefont {G.}~\bibnamefont
  {Bencedi}} (\bibinfo {collaboration} {ALICE}),\ }in\ \href@noop {} {\emph
  {\bibinfo {booktitle} {{Proceedings, 51st Rencontres de Moriond on QCD and
  High Energy Interactions: La Thuile, Italy, March 19-26, 2016}}}}\ (\bibinfo
  {year} {2016})\ pp.\ \bibinfo {pages} {285--288},\ \Eprint
  {http://arxiv.org/abs/1609.05665} {arXiv:1609.05665 [nucl-ex]} \BibitemShut
  {NoStop}%
\bibitem [{\citenamefont {Sett}(2017)}]{Sett:2015pba}%
  \BibitemOpen
  \bibfield  {author} {\bibinfo {author} {\bibfnamefont {P.}~\bibnamefont
  {Sett}} (\bibinfo {collaboration} {PHENIX}),\ }\bibfield  {booktitle} {\emph
  {\bibinfo {booktitle} {{Proceedings, 7th International Conference on Physics
  and Astrophysics of Quark Gluon Plasma (ICPAQGP 2015): Kolkata, West Bengal,
  India, February 2-6, 2015}}},\ }\href {\doibase 10.22323/1.242.0091}
  {\bibfield  {journal} {\bibinfo  {journal} {PoS}\ }\textbf {\bibinfo {volume}
  {ICPAQGP2015}},\ \bibinfo {pages} {091} (\bibinfo {year} {2017})}\BibitemShut
  {NoStop}%
\bibitem [{\citenamefont {Todoroki}(2017)}]{Todoroki:2017ngs}%
  \BibitemOpen
  \bibfield  {author} {\bibinfo {author} {\bibfnamefont {T.}~\bibnamefont
  {Todoroki}} (\bibinfo {collaboration} {STAR}),\ }\bibfield  {booktitle}
  {\emph {\bibinfo {booktitle} {{Proceedings, 26th International Conference on
  Ultra-relativistic Nucleus-Nucleus Collisions (Quark Matter 2017): Chicago,
  Illinois, USA, February 5-11, 2017}}},\ }\href {\doibase
  10.1016/j.nuclphysa.2017.04.020} {\bibfield  {journal} {\bibinfo  {journal}
  {Nucl. Phys.}\ }\textbf {\bibinfo {volume} {A967}},\ \bibinfo {pages} {572}
  (\bibinfo {year} {2017})},\ \Eprint {http://arxiv.org/abs/1704.06251}
  {arXiv:1704.06251 [nucl-ex]} \BibitemShut {NoStop}%
\bibitem [{\citenamefont {Wang}(2001)}]{wang:2000fq}%
  \BibitemOpen
  \bibfield  {author} {\bibinfo {author} {\bibfnamefont {X.-N.}\ \bibnamefont
  {Wang}},\ }\href {\doibase 10.1103/PhysRevC.63.054902} {\bibfield  {journal}
  {\bibinfo  {journal} {Phys. Rev.}\ }\textbf {\bibinfo {volume} {C63}},\
  \bibinfo {pages} {054902} (\bibinfo {year} {2001})},\ \Eprint
  {http://arxiv.org/abs/nucl-th/0009019} {arXiv:nucl-th/0009019 [nucl-th]}
  \BibitemShut {NoStop}%
\bibitem [{\citenamefont {Gyulassy}\ \emph {et~al.}(2001)\citenamefont
  {Gyulassy}, \citenamefont {Vitev},\ and\ \citenamefont
  {Wang}}]{gyulassy:2000gk}%
  \BibitemOpen
  \bibfield  {author} {\bibinfo {author} {\bibfnamefont {M.}~\bibnamefont
  {Gyulassy}}, \bibinfo {author} {\bibfnamefont {I.}~\bibnamefont {Vitev}}, \
  and\ \bibinfo {author} {\bibfnamefont {X.~N.}\ \bibnamefont {Wang}},\ }\href
  {\doibase 10.1103/PhysRevLett.86.2537} {\bibfield  {journal} {\bibinfo
  {journal} {Phys. Rev. Lett.}\ }\textbf {\bibinfo {volume} {86}},\ \bibinfo
  {pages} {2537} (\bibinfo {year} {2001})},\ \Eprint
  {http://arxiv.org/abs/nucl-th/0012092} {arXiv:nucl-th/0012092 [nucl-th]}
  \BibitemShut {NoStop}%
\bibitem [{\citenamefont {Liao}\ and\ \citenamefont
  {Shuryak}(2009)}]{Liao:2008dk}%
  \BibitemOpen
  \bibfield  {author} {\bibinfo {author} {\bibfnamefont {J.}~\bibnamefont
  {Liao}}\ and\ \bibinfo {author} {\bibfnamefont {E.}~\bibnamefont {Shuryak}},\
  }\href {\doibase 10.1103/PhysRevLett.102.202302} {\bibfield  {journal}
  {\bibinfo  {journal} {Phys. Rev. Lett.}\ }\textbf {\bibinfo {volume} {102}},\
  \bibinfo {pages} {202302} (\bibinfo {year} {2009})},\ \Eprint
  {http://arxiv.org/abs/0810.4116} {arXiv:0810.4116 [nucl-th]} \BibitemShut
  {NoStop}%
\bibitem [{\citenamefont {Jia}\ \emph {et~al.}(2011)\citenamefont {Jia},
  \citenamefont {Horowitz},\ and\ \citenamefont {Liao}}]{Jia:2011pi}%
  \BibitemOpen
  \bibfield  {author} {\bibinfo {author} {\bibfnamefont {J.}~\bibnamefont
  {Jia}}, \bibinfo {author} {\bibfnamefont {W.~A.}\ \bibnamefont {Horowitz}}, \
  and\ \bibinfo {author} {\bibfnamefont {J.}~\bibnamefont {Liao}},\ }\href
  {\doibase 10.1103/PhysRevC.84.034904} {\bibfield  {journal} {\bibinfo
  {journal} {Phys. Rev.}\ }\textbf {\bibinfo {volume} {C84}},\ \bibinfo {pages}
  {034904} (\bibinfo {year} {2011})},\ \Eprint {http://arxiv.org/abs/1101.0290}
  {arXiv:1101.0290 [nucl-th]} \BibitemShut {NoStop}%
\bibitem [{\citenamefont {Zapp}(2014)}]{Zapp:2013zya}%
  \BibitemOpen
  \bibfield  {author} {\bibinfo {author} {\bibfnamefont {K.~C.}\ \bibnamefont
  {Zapp}},\ }\href {\doibase 10.1016/j.physletb.2014.06.020} {\bibfield
  {journal} {\bibinfo  {journal} {Phys. Lett.}\ }\textbf {\bibinfo {volume}
  {B735}},\ \bibinfo {pages} {157} (\bibinfo {year} {2014})},\ \Eprint
  {http://arxiv.org/abs/1312.5536} {arXiv:1312.5536 [hep-ph]} \BibitemShut
  {NoStop}%
\bibitem [{\citenamefont {Betz}\ and\ \citenamefont
  {Gyulassy}(2014)}]{betz:2014cza}%
  \BibitemOpen
  \bibfield  {author} {\bibinfo {author} {\bibfnamefont {B.}~\bibnamefont
  {Betz}}\ and\ \bibinfo {author} {\bibfnamefont {M.}~\bibnamefont
  {Gyulassy}},\ }\href {\doibase 10.1007/JHEP10(2014)043,
  10.1007/JHEP08(2014)090} {\bibfield  {journal} {\bibinfo  {journal} {JHEP}\
  }\textbf {\bibinfo {volume} {08}},\ \bibinfo {pages} {090} (\bibinfo {year}
  {2014})},\ \bibinfo {note} {[Erratum: JHEP10,043(2014)]},\ \Eprint
  {http://arxiv.org/abs/1404.6378} {arXiv:1404.6378 [hep-ph]} \BibitemShut
  {NoStop}%
\bibitem [{\citenamefont {Nahrgang}\ \emph {et~al.}(2015)\citenamefont
  {Nahrgang}, \citenamefont {Aichelin}, \citenamefont {Bass}, \citenamefont
  {Gossiaux},\ and\ \citenamefont {Werner}}]{Nahrgang:2014vza}%
  \BibitemOpen
  \bibfield  {author} {\bibinfo {author} {\bibfnamefont {M.}~\bibnamefont
  {Nahrgang}}, \bibinfo {author} {\bibfnamefont {J.}~\bibnamefont {Aichelin}},
  \bibinfo {author} {\bibfnamefont {S.}~\bibnamefont {Bass}}, \bibinfo {author}
  {\bibfnamefont {P.~B.}\ \bibnamefont {Gossiaux}}, \ and\ \bibinfo {author}
  {\bibfnamefont {K.}~\bibnamefont {Werner}},\ }\href {\doibase
  10.1103/PhysRevC.91.014904} {\bibfield  {journal} {\bibinfo  {journal} {Phys.
  Rev.}\ }\textbf {\bibinfo {volume} {C91}},\ \bibinfo {pages} {014904}
  (\bibinfo {year} {2015})},\ \Eprint {http://arxiv.org/abs/1410.5396}
  {arXiv:1410.5396 [hep-ph]} \BibitemShut {NoStop}%
\bibitem [{\citenamefont {Xu}\ \emph {et~al.}(2015)\citenamefont {Xu},
  \citenamefont {Liao},\ and\ \citenamefont {Gyulassy}}]{Xu:2014tda}%
  \BibitemOpen
  \bibfield  {author} {\bibinfo {author} {\bibfnamefont {J.}~\bibnamefont
  {Xu}}, \bibinfo {author} {\bibfnamefont {J.}~\bibnamefont {Liao}}, \ and\
  \bibinfo {author} {\bibfnamefont {M.}~\bibnamefont {Gyulassy}},\ }\href
  {\doibase 10.1088/0256-307X/32/9/092501} {\bibfield  {journal} {\bibinfo
  {journal} {Chin. Phys. Lett.}\ }\textbf {\bibinfo {volume} {32}},\ \bibinfo
  {pages} {092501} (\bibinfo {year} {2015})},\ \Eprint
  {http://arxiv.org/abs/1411.3673} {arXiv:1411.3673 [hep-ph]} \BibitemShut
  {NoStop}%
\bibitem [{\citenamefont {Xu}\ \emph {et~al.}(2018)\citenamefont {Xu},
  \citenamefont {Bernhard}, \citenamefont {Bass}, \citenamefont {Nahrgang},\
  and\ \citenamefont {Cao}}]{Xu:2017obm}%
  \BibitemOpen
  \bibfield  {author} {\bibinfo {author} {\bibfnamefont {Y.}~\bibnamefont
  {Xu}}, \bibinfo {author} {\bibfnamefont {J.~E.}\ \bibnamefont {Bernhard}},
  \bibinfo {author} {\bibfnamefont {S.~A.}\ \bibnamefont {Bass}}, \bibinfo
  {author} {\bibfnamefont {M.}~\bibnamefont {Nahrgang}}, \ and\ \bibinfo
  {author} {\bibfnamefont {S.}~\bibnamefont {Cao}},\ }\href {\doibase
  10.1103/PhysRevC.97.014907} {\bibfield  {journal} {\bibinfo  {journal} {Phys.
  Rev.}\ }\textbf {\bibinfo {volume} {C97}},\ \bibinfo {pages} {014907}
  (\bibinfo {year} {2018})},\ \Eprint {http://arxiv.org/abs/1710.00807}
  {arXiv:1710.00807 [nucl-th]} \BibitemShut {NoStop}%
\bibitem [{\citenamefont {Cao}\ \emph {et~al.}(2019)\citenamefont {Cao} \emph
  {et~al.}}]{Cao:2018ews}%
  \BibitemOpen
  \bibfield  {author} {\bibinfo {author} {\bibfnamefont {S.}~\bibnamefont
  {Cao}} \emph {et~al.},\ }\href {\doibase 10.1103/PhysRevC.99.054907}
  {\bibfield  {journal} {\bibinfo  {journal} {Phys. Rev.}\ }\textbf {\bibinfo
  {volume} {C99}},\ \bibinfo {pages} {054907} (\bibinfo {year} {2019})},\
  \Eprint {http://arxiv.org/abs/1809.07894} {arXiv:1809.07894 [nucl-th]}
  \BibitemShut {NoStop}%
\bibitem [{\citenamefont {Sun}\ \emph {et~al.}(2019)\citenamefont {Sun},
  \citenamefont {Coci}, \citenamefont {Das}, \citenamefont {Plumari},
  \citenamefont {Ruggieri},\ and\ \citenamefont {Greco}}]{Sun:2019fud}%
  \BibitemOpen
  \bibfield  {author} {\bibinfo {author} {\bibfnamefont {Y.}~\bibnamefont
  {Sun}}, \bibinfo {author} {\bibfnamefont {G.}~\bibnamefont {Coci}}, \bibinfo
  {author} {\bibfnamefont {S.~K.}\ \bibnamefont {Das}}, \bibinfo {author}
  {\bibfnamefont {S.}~\bibnamefont {Plumari}}, \bibinfo {author} {\bibfnamefont
  {M.}~\bibnamefont {Ruggieri}}, \ and\ \bibinfo {author} {\bibfnamefont
  {V.}~\bibnamefont {Greco}},\ }\href@noop {} {\  (\bibinfo {year} {2019})},\
  \Eprint {http://arxiv.org/abs/1902.06254} {arXiv:1902.06254 [nucl-th]}
  \BibitemShut {NoStop}%
\bibitem [{\citenamefont {Broniowski}\ and\ \citenamefont
  {Ruiz~Arriola}(2014)}]{Broniowski:2013dia}%
  \BibitemOpen
  \bibfield  {author} {\bibinfo {author} {\bibfnamefont {W.}~\bibnamefont
  {Broniowski}}\ and\ \bibinfo {author} {\bibfnamefont {E.}~\bibnamefont
  {Ruiz~Arriola}},\ }\href {\doibase 10.1103/PhysRevLett.112.112501} {\bibfield
   {journal} {\bibinfo  {journal} {Phys. Rev. Lett.}\ }\textbf {\bibinfo
  {volume} {112}},\ \bibinfo {pages} {112501} (\bibinfo {year} {2014})},\
  \Eprint {http://arxiv.org/abs/1312.0289} {arXiv:1312.0289 [nucl-th]}
  \BibitemShut {NoStop}%
\bibitem [{\citenamefont {Adamczyk}\ \emph {et~al.}(2015)\citenamefont
  {Adamczyk} \emph {et~al.}}]{Adamczyk:2015obl}%
  \BibitemOpen
  \bibfield  {author} {\bibinfo {author} {\bibfnamefont {L.}~\bibnamefont
  {Adamczyk}} \emph {et~al.} (\bibinfo {collaboration} {STAR}),\ }\href
  {\doibase 10.1103/PhysRevLett.115.222301} {\bibfield  {journal} {\bibinfo
  {journal} {Phys. Rev. Lett.}\ }\textbf {\bibinfo {volume} {115}},\ \bibinfo
  {pages} {222301} (\bibinfo {year} {2015})},\ \Eprint
  {http://arxiv.org/abs/1505.07812} {arXiv:1505.07812 [nucl-ex]} \BibitemShut
  {NoStop}%
\bibitem [{\citenamefont {Wang}\ and\ \citenamefont
  {Sorensen}(2014)}]{wang:2014qxa}%
  \BibitemOpen
  \bibfield  {author} {\bibinfo {author} {\bibfnamefont {H.}~\bibnamefont
  {Wang}}\ and\ \bibinfo {author} {\bibfnamefont {P.}~\bibnamefont {Sorensen}}
  (\bibinfo {collaboration} {STAR}),\ }\bibfield  {booktitle} {\emph {\bibinfo
  {booktitle} {{Proceedings, 6th International Conference on Hard and
  Electromagnetic Probes of High-Energy Nuclear Collisions (Hard Probes 2013):
  Cape Town, South Africa, November 4-8, 2013}}},\ }\href {\doibase
  10.1016/j.nuclphysa.2014.09.111} {\bibfield  {journal} {\bibinfo  {journal}
  {Nucl. Phys.}\ }\textbf {\bibinfo {volume} {A932}},\ \bibinfo {pages} {169}
  (\bibinfo {year} {2014})},\ \Eprint {http://arxiv.org/abs/1406.7522}
  {arXiv:1406.7522 [nucl-ex]} \BibitemShut {NoStop}%
\bibitem [{\citenamefont {Goldschmidt}\ \emph {et~al.}(2015)\citenamefont
  {Goldschmidt}, \citenamefont {Qiu}, \citenamefont {Shen},\ and\ \citenamefont
  {Heinz}}]{Goldschmidt:2015kpa}%
  \BibitemOpen
  \bibfield  {author} {\bibinfo {author} {\bibfnamefont {A.}~\bibnamefont
  {Goldschmidt}}, \bibinfo {author} {\bibfnamefont {Z.}~\bibnamefont {Qiu}},
  \bibinfo {author} {\bibfnamefont {C.}~\bibnamefont {Shen}}, \ and\ \bibinfo
  {author} {\bibfnamefont {U.}~\bibnamefont {Heinz}},\ }\href {\doibase
  10.1103/PhysRevC.92.044903} {\bibfield  {journal} {\bibinfo  {journal} {Phys.
  Rev.}\ }\textbf {\bibinfo {volume} {C92}},\ \bibinfo {pages} {044903}
  (\bibinfo {year} {2015})},\ \Eprint {http://arxiv.org/abs/1507.03910}
  {arXiv:1507.03910 [nucl-th]} \BibitemShut {NoStop}%
\bibitem [{\citenamefont {Rybczyński}\ \emph {et~al.}(2018)\citenamefont
  {Rybczyński}, \citenamefont {Piotrowska},\ and\ \citenamefont
  {Broniowski}}]{Rybczynski:2017nrx}%
  \BibitemOpen
  \bibfield  {author} {\bibinfo {author} {\bibfnamefont {M.}~\bibnamefont
  {Rybczyński}}, \bibinfo {author} {\bibfnamefont {M.}~\bibnamefont
  {Piotrowska}}, \ and\ \bibinfo {author} {\bibfnamefont {W.}~\bibnamefont
  {Broniowski}},\ }\href {\doibase 10.1103/PhysRevC.97.034912} {\bibfield
  {journal} {\bibinfo  {journal} {Phys. Rev.}\ }\textbf {\bibinfo {volume}
  {C97}},\ \bibinfo {pages} {034912} (\bibinfo {year} {2018})},\ \Eprint
  {http://arxiv.org/abs/1711.00438} {arXiv:1711.00438 [nucl-th]} \BibitemShut
  {NoStop}%
\bibitem [{\citenamefont {Schenke}\ \emph {et~al.}(2019)\citenamefont
  {Schenke}, \citenamefont {Shen},\ and\ \citenamefont
  {Tribedy}}]{Schenke:2019ruo}%
  \BibitemOpen
  \bibfield  {author} {\bibinfo {author} {\bibfnamefont {B.}~\bibnamefont
  {Schenke}}, \bibinfo {author} {\bibfnamefont {C.}~\bibnamefont {Shen}}, \
  and\ \bibinfo {author} {\bibfnamefont {P.}~\bibnamefont {Tribedy}},\
  }\href@noop {} {\  (\bibinfo {year} {2019})},\ \Eprint
  {http://arxiv.org/abs/1901.04378} {arXiv:1901.04378 [nucl-th]} \BibitemShut
  {NoStop}%
\bibitem [{\citenamefont {Collaboration}(2018)}]{CMS:2018jmx}%
  \BibitemOpen
  \bibfield  {author} {\bibinfo {author} {\bibfnamefont {C.}~\bibnamefont
  {Collaboration}} (\bibinfo {collaboration} {CMS}),\ }\href@noop {} {\
  (\bibinfo {year} {2018})}\BibitemShut {NoStop}%
\bibitem [{\citenamefont {Acharya}\ \emph {et~al.}(2018)\citenamefont {Acharya}
  \emph {et~al.}}]{Acharya:2018ihu}%
  \BibitemOpen
  \bibfield  {author} {\bibinfo {author} {\bibfnamefont {S.}~\bibnamefont
  {Acharya}} \emph {et~al.} (\bibinfo {collaboration} {ALICE}),\ }\href
  {\doibase 10.1016/j.physletb.2018.06.059} {\bibfield  {journal} {\bibinfo
  {journal} {Phys. Lett.}\ }\textbf {\bibinfo {volume} {B784}},\ \bibinfo
  {pages} {82} (\bibinfo {year} {2018})},\ \Eprint
  {http://arxiv.org/abs/1805.01832} {arXiv:1805.01832 [nucl-ex]} \BibitemShut
  {NoStop}%
\bibitem [{\citenamefont {collaboration}(2018)}]{ATLAS:2018iom}%
  \BibitemOpen
  \bibfield  {author} {\bibinfo {author} {\bibfnamefont {T.~A.}\ \bibnamefont
  {collaboration}} (\bibinfo {collaboration} {ATLAS}),\ }\bibinfo
  {organization} {CERN}\ (\bibinfo  {publisher} {CERN},\ \bibinfo {address}
  {Geneva},\ \bibinfo {year} {2018})\BibitemShut {NoStop}%
\bibitem [{\citenamefont {Plumari}\ \emph {et~al.}(2019)\citenamefont
  {Plumari}, \citenamefont {Coci}, \citenamefont {Das}, \citenamefont
  {Minissale},\ and\ \citenamefont {Greco}}]{Plumari:2019yhg}%
  \BibitemOpen
  \bibfield  {author} {\bibinfo {author} {\bibfnamefont {S.}~\bibnamefont
  {Plumari}}, \bibinfo {author} {\bibfnamefont {G.}~\bibnamefont {Coci}},
  \bibinfo {author} {\bibfnamefont {S.~K.}\ \bibnamefont {Das}}, \bibinfo
  {author} {\bibfnamefont {V.}~\bibnamefont {Minissale}}, \ and\ \bibinfo
  {author} {\bibfnamefont {V.}~\bibnamefont {Greco}},\ }\bibfield  {booktitle}
  {\emph {\bibinfo {booktitle} {{Proceedings, 27th International Conference on
  Ultrarelativistic Nucleus-Nucleus Collisions (Quark Matter 2018): Venice,
  Italy, May 14-19, 2018}}},\ }\href {\doibase 10.1016/j.nuclphysa.2018.10.070}
  {\bibfield  {journal} {\bibinfo  {journal} {Nucl. Phys.}\ }\textbf {\bibinfo
  {volume} {A982}},\ \bibinfo {pages} {655} (\bibinfo {year} {2019})},\ \Eprint
  {http://arxiv.org/abs/1901.07815} {arXiv:1901.07815 [hep-ph]} \BibitemShut
  {NoStop}%
\end{thebibliography}%
\end{document}